\documentclass[]{aastex631}
\usepackage{amsmath,acronym}
\hypersetup{linkcolor=red,citecolor=blue,filecolor=cyan,urlcolor=magenta}
\newcommand{\Mpc}{\,{\rm Mpc}}
\newcommand{\Msun}{\,{\rm M}_\odot}

\newcommand{\fss}{\mathcal{F}{\rm -statistic}}

\newcommand{\PKU}{Kavli Institute for Astronomy and Astrophysics, Peking University, Beijing 100871, People's Republic of China}
\newcommand{\PKUA}{Department of Astronomy, School of Physics, Peking University, Beijing 100871, People's Republic of China}
\newcommand{\NAOCAS}{National Astronomical Observatories, Chinese Academy of Sciences, Beijing 100012, People's Republic of China}
\def\apjl{Astrophysical Journal, Letters}             % Astrophysical Journal, Letters
\def\apjs{ApJS}              % Astrophysical Journal, Supplement
\def\aap{A\&A}                % Astronomy and Astrophysics
\acrodef{GW}{gravitational wave}
\acrodef{GWTC-3}{the third Gravitational-wave Transient Catalog}
\acrodef{FFT}{fast Fourier transform}
\acrodef{LIGO}{Laser Interferometer Gravitational-Wave Observatory}
\acrodef{LVC}{LIGO-Virgo Collaboration}
\acrodef{LVK}{LIGO-Virgo-KAGRA}
\acrodef{BNS}{binary neutron star}
\acrodef{NR}{numerical relativity}
\acrodef{TD}{time-domain}
\acrodef{ET}{Einstein Telescope}
\acrodef{BH}{black hole}
\acrodef{PE}{parameter estimation}
\acrodef{BBH}{binary black hole}
\acrodef{GR}{general relativity}
\acrodef{PSD}{power spectral density}
\acrodef{PDF}{probability density function}
\acrodef{ACF}{autocovariance function}
\acrodef{QNMs}{quasinormal modes}
\acrodef{SNR}{signal-to-noise ratio}

\shorttitle{Ringdown analysis using the $\mathcal{F}$-statistic}
\begin{document}

\title{Gravitational wave ringdown analysis using the $\mathcal{F}$-statistic}

\correspondingauthor{Hai-Tian Wang, Lijing Shao}
\email{wanght@pku.edu.cn, lshao@pku.edu.cn}
\author[0000-0002-7779-8239]{\textcolor{blue}{Hai-Tian Wang}}
\affiliation{\PKU}
\author[0000-0001-8548-9535]{\textcolor{blue}{Garvin Yim}}
\affiliation{\PKU}
\author[0000-0003-3950-9317]{\textcolor{blue}{Xian Chen}}
\affiliation{\PKUA}
\affiliation{\PKU}
\author[0000-0002-1334-8853]{\textcolor{blue}{Lijing Shao}}
\affiliation{\PKU}
\affiliation{\NAOCAS}

\begin{abstract}
After the final stage of the merger of two black holes, the ringdown signal takes an important role on providing information about the gravitational dynamics in strong field.
We introduce a novel time-domain (TD) approach, predicated on the $\mathcal{F}$-statistic, for ringdown analysis.
This method diverges from traditional TD techniques in that its parameter space remains constant irrespective of the number of modes incorporated. 
This feature is achieved by reconfiguring the likelihood and analytically maximizing over the extrinsic parameters that encompass the amplitudes and reference phases of all modes. 
Consequently, when performing the ringdown analysis under the assumption that the ringdown signal is detected by the Einstein Telescope, parameter estimation computation time is shortened by at most five orders of magnitude compared to the traditional TD method. 
We further establish that traditional TD methods become difficult when including multiple overtone modes due to close oscillation frequencies and damping times across different overtone modes. 
Encouragingly, this issue is effectively addressed by our new TD technique. 
The accessibility of this new TD method extends to a broad spectrum of research and offers flexibility for various topics within black hole spectroscopy applicable to both current and future gravitational wave detectors.
\end{abstract}


%%%%%%%%%%%%%%%%%%%%%%%%%%%%%%%%%%%%%%%%%%%%%%%%%%%%%%%%%%%%%%%%
\section{Introduction}\label{sec:intro}
%% QNM
According to \ac{GR}, the \ac{GW} signal from the ringdown of a \ac{BH} is characterized by the amalgamation of \ac{QNMs} \citep{Schw_PRD_Vishveshwara1970,GW_APJL_Press1971,QNM_APJ_Teukolsky1973}, which can further be decomposed into spin-weighted spheroidal harmonics with angular indices $(\ell,m)$. Each set of these angular indices encompasses a series of overtone modes denoted by $n$ \citep{Berti:2009kk}. Research focused on extracting information from these modes is referred to as ``\ac{BH} spectroscopy" \citep{Dreyer:2003bv,Berti:2005ys,Berti:2016lat,Yang:2017zxs,Isi:2019aib,Bhagwat:2019dtm,Ma:2023cwe}.

%% overtone modes
Typically, overtone modes exhibit a more rapid decay than the fundamental mode $(\ell=m=2, \, n=0)$ and higher multipoles. The latter are postulated to be significant for systems with asymmetric mass ratios \citep{Berti:2007zu,Gossan:2011ha,Brito:2018rfr}. Investigations \citep{Capano:2021etf,Capano:2022zqm,Abedi:2023kot,Siegel:2023lxl} have identified evidence of various higher multipoles from the ringdown analysis of GW190521 \citep{2020PhRvL.125j1102A}, an event potentially characterized by an asymmetric mass ratio \citep{Estelles:2021jnz,Nitz:2020mga}. Early research largely overlooked the contribution of overtone modes \citep{Berti:2005ys,Gossan:2011ha} until it was discovered by \citet{Overtone_PRX_Giesler2019} that when $7$ overtone modes are incorporated, the ringdown waveform aligns with the peak amplitude of \ac{NR} waveforms. However, these overtone modes contribute minimally to the GW strain. Despite there being numerous \ac{GW} events detected by the \ac{LVK} Collaboration \citep{LIGO_PRX2019,LIGO_O3a_PRX2020,KAGRA:2021vkt}, only weak evidence has been found for the first overtone mode, even when matched from the peak amplitude \citep{2021arXiv211206861T}. Vigorous debates continue on this topic from the inaugural \ac{GW} event, GW150914 \citep{Isi:2019aib,2021PhRvD.103l2002A,2021arXiv211206861T,Isi:2023nif,Carullo:2023gtf}, including from a subset of us where we showed that there was only very weak evidence for the first overtone mode in GW150914 \citep{Wang:2023mst}. This conclusion was reached through the use of a carefully verified noise estimation method \citep{Wang:2024liy}.
Besides this, there are also some studies \citep{Baibhav:2023clw,Nee:2023osy,Zhu:2023mzv} which argue that higher overtones $(n>2)$ overfit the transient radiation and nonlinearities close to the merger.
It is crucial to validate theoretical assertions of this by analyzing real \ac{GW} data or meticulously simulated \ac{GW} data.

%% motivation
However, two factors currently hinder prospective \ac{PE} studies which include multiple overtone modes. The first factor is the proximity of oscillation frequencies and damping times across different overtone modes \citep{Cabero:2019zyt,Maselli:2019mjd}, rendering them nearly indistinguishable. The second factor is the expansion of parameter space when more overtone modes are incorporated, despite there being only a limited increase in the \ac{SNR} in contribution to the strain. For instance, in the ringdown waveform examined by \citet{Bhagwat:2019dtm}, each additional overtone mode introduces four independent parameters. Consequently, with $8$ overtones included in the ringdown waveform, there would be at least $32$ parameters---a situation ``{\it which makes performing Bayesian \ac{PE} infeasible}" \citep{Bhagwat:2019dtm}.

%% F-statistic
To address these challenges, we propose a method that integrates the $\mathcal{F}$-statistic with the traditional \ac{TD} (TTD) method \citep{2021arXiv210705609I}. The $\mathcal{F}$-statistic approach was initially formulated for continuous \ac{GW} signals \citep{Jaranowski:1998qm,Cutler:2005hc,Dreissigacker:2018afk} and later applied to extreme mass-ratio inspiral signals \citep{Wang:2012xh}. A shared characteristic of these signal types is their non-smooth spectrum featuring multiple peaks. Consequently, the likelihood hyper-surface contains numerous close local maxima \citep{Babak:2014kqa}, like a forest, which traditional Bayesian inference struggles to resolve effectively. The $\mathcal{F}$-statistic aids in reducing the parameter space by analytically maximizing over all extrinsic parameters, thereby enhancing efficiency in \ac{PE}. 
This technique has found extensive applications in continuous \ac{GW} searches \citep{LIGOScientific:2010tql,LIGOScientific:2019yhl,Sieniawska:2019hmd,LIGOScientific:2021mwx,Steltner:2023cfk,Wette:2023dom}.

%% usage
In the context of \ac{PE} for ringdown signals, we encounter analogous challenges in having additional parameters. Our investigation, using the $\mathcal{F}$-statistic, reveals that each overtone mode introduces only two additional parameters. This implies that the $\mathcal{F}$-statistic enhances efficiency in the \ac{PE} of ringdown. 
Unless otherwise stated, we employ geometric units with $G=c=1$.

%% ------------------------
\section{Formulating the $\mathcal{F}$-statistic}\label{sec:Fs}
%% RD waveform
The TD ringdown waveform of a Kerr \ac{BH} can be expressed as 
\begin{equation}
\begin{aligned}
\label{eq:rin_td}
h_{+}(t)+ih_{\times}(t)
=\sum_{\ell}\sum_m\sum^N_n{}_{-2}S_{\ell m}(\iota, \delta)A_{\ell mn}
\exp\left(i2\pi f_{\ell mn}t+i\phi_{\ell mn}-\frac{t}{\tau_{\ell mn}}\right).
\end{aligned}
\end{equation}
In this equation, $N$ signifies the total number of modes, including the fundamental mode and overtone modes, each labelled by $n = 0, 1, \cdots$. 
The variables $A_{\ell mn}$ and $\phi_{\ell mn}$ correspond to the amplitudes and phases for each mode, respectively. The inclination and azimuthal angles are represented by $\iota$ and $\delta$, with the latter being set to zero for our investigation. 
The (real) oscillation frequency is denoted by $f_{\ell mn}$, while $\tau_{\ell mn}$ represents the damping time; both quantities are determined by the final mass ($M_f$) and final spin ($\chi_f$) of the remnant \ac{BH}. Thus, in our study, each overtone mode introduces two more parameters, which is different from that in \citet{Bhagwat:2019dtm}. Finally, $_{-2}S_{\ell m}$ represents the spin-weighted spheroidal harmonics \citep{QNM_APJ_Teukolsky1973}, which we approximate as spin-weighted spherical harmonics for reasons detailed by \citet{Overtone_PRX_Giesler2019}. 
A list of spin-weighted spherical harmonics can be found in \citet{brugmann_PRD2008} with $s=-2$. 

%% signal of ET
In the pursuit of discerning multiple overtone modes, we employ a next generation ground-based detector, the \ac{ET} in its ET-D configuration \citep{2010CQGra..27s4002P, Hild:2010id}. Conventionally, the \ac{GW} signal identified by such a detector is expressed as $h(t)=F^+h_++F^{\times}h_{\times}$, where $F^{+,\times}$ denotes the antenna pattern functions that are contingent on both sky location and the \ac{GW} polarization angle. Each mode present in Eq.~(\ref{eq:rin_td}) can be reformulated into a $B^{\ell mn,k}h_{\ell mn,k}$ form, with $k=1,2$ and 
\begin{equation}
\begin{aligned}
B^{\ell mn,1}=&A_{\ell mn}\cos\phi_{\ell mn},\\
B^{\ell mn,2}=&A_{\ell mn}\sin\phi_{\ell mn},\\
h_{\ell mn,1}=&\left[F^+\cos(2\pi f_{\ell mn}t)+F^{\times}\sin(2\pi f_{\ell mn}t)\right]
{}_{-2}Y_{\ell m}(\iota, \delta)\exp \Big(-\frac{t}{\tau_{\ell mn}} \Big),\\
h_{\ell mn,2}=&\left[-F^{+}\sin(2\pi f_{\ell mn}t)+F^{\times}\cos(2\pi f_{\ell mn}t)\right]
{}_{-2}Y_{\ell m}(\iota,\delta)\exp \Big(-\frac{t}{\tau_{\ell mn}} \Big).
\label{eq:rin_split}
\end{aligned}
\end{equation}
As seen in Eq.~(\ref{eq:rin_split}), for each mode, $B^{\ell mn,k}$ is solely dependent on two extrinsic parameters, $A_{\ell mn}$ and $\phi_{\ell mn}$.
After the reformulation, the ringdown signal can be written as $h(t)=B^{\mu}h_{\mu}$, where $\mu=\{(220,1),(220,2),\ldots,(\ell mn,1),(\ell mn,2)\}$ and the length of $\mu$ is $2\times N$.

%% simplify the APF
In the present study, we consider the sky-averaged antenna pattern functions, resulting in $\langle F^2_+\rangle=\langle F^2_{\times}\rangle=\sin^2\zeta/5$ for a detector with an arm opening angle $\zeta$ \citep{Jaranowski:1998qm}. Once built, ET will be composed of three detectors and have $\zeta = \pi/3$. Consequently, $F^{+,\times} = \sqrt{15}/10$ and the detected signal can be represented as $h(t)=\sqrt{15}/10(h_++h_{\times})$. 
It should be noted that using the average beam patterns may lead to some bias in the final mass and spin. However, future analyses can straightforwardly incorporate source and detector positions and orientations to address this \citep{Jaranowski:1998qm}. As this is the first work to implement the $\fss$ in ringdown analyses, our primary focus is on assessing the efficiency of this method compared to the TTD method, so we assume the sky-averaged case for both methods, allowing for a fair and simple comparison.

%% 0305
In order to emulate \ac{GW} data, a GW150914-like \ac{NR} waveform, SXS:BBH:0305, is incorporated into the noise of ET. This particular waveform is part of the Simulation eXtreme Spacetimes catalog \citep{Boyle:2019kee}, and characterizes a non-precessing source with a mass ratio of $0.82$ and a remnant possessing a dimensionless spin of $0.69$. A luminosity distance of $390\Mpc$, an inclination angle of $3\pi/4$, and a reference phase of $0$ are utilized in this study. Assuming that the redshifted chirp mass equates to $31\Msun$, it follows that the redshifted final mass is approximately $68.2\Msun$. The focus here lies solely on the ringdown signal from the multipole where $\ell=|m|=2$, with $h_{\ell m}=(-1)^{\ell}h^*_{\ell-m}$. Mode-mixing contributions are not taken into account, aligning with \citet{Overtone_PRX_Giesler2019}. Consequently, Eq.~(\ref{eq:rin_split}) should be modified by substituting ${}_{-2}Y_{\ell m}F^+$ and ${}_{-2}Y_{\ell m}F^{\times}$ with their respective counterparts, namely $[{}_{-2}Y_{\ell m}+(-1)^{\ell}{}_{-2}Y_{\ell -m}]F^+$ and $[{}_{-2}Y_{\ell m}-(-1)^{\ell}{}_{-2}Y_{\ell-m}]F^\times$. 
The incorporation of higher overtone modes does not imply that we subscribe to the notion that the post-peak signal can be linearly accounted for by these modes.
As illustrated in Sec.~\ref{sec:intro}, our primary objective centers on devising a novel method to undertake these pivotal analyses with upcoming \ac{GW} datasets.
In this context, we employ scenarios involving multiple overtone modes to demonstrate the efficacy of this innovative approach.
Although overtone modes serve as the primary example for assessing the efficacy of the $\mathcal{F}$-statistic, the implementation can be readily extended to other scenarios such as different types of \ac{QNMs}.

%% noise data
The noise data, derived from the ET-D noise curve \citep{Hild:2010id}, is assumed to be Gaussian and stationary. As such, it is described by a multivariate normal distribution $\vec{n}\sim\mathcal{N}(\vec{0},\mathcal{C})$, where $\mathcal{C}$ represents the covariance matrix, which is provided by the auto-covariance function. Utilizing the Wiener-Khinchin theorem allows for the extraction of the auto-covariance function from the one-sided \ac{PSD}. This is achieved through the application of Welch's method to the noise data \citep{1967D.Welch}.
In this case, the \ac{SNR} of the ringdown signal is approximately $312$, calculated under the assumption that it starts from the peak amplitude.

%% log-likelihood
In the process of extracting ringdown parameters from discrete \ac{GW} data $\vec{d}$, we employ an algorithm that is fundamentally based on the Bayes' theorem. It is expressed as $P(\theta|\vec{d},I)=P(\vec{d}|\theta,I)P(\theta|I)/P(\vec{d}|I)$, where $P(\theta|\vec{d},I)$ represents the desired posterior, $P(\vec{d}|\theta,I)$ signifies the likelihood function, and $P(\theta|I)$ denotes the prior. Additionally, $P(\vec{d}|I)$ represents the evidence while $\theta$ represents the model parameters and finally, $I$ indicates other background knowledge of a selected model. 
In the \ac{TD}, the log-likelihood function can be expressed as
\begin{equation}
\begin{aligned}
	\ln \mathcal{L}&=-\frac{1}{2}\big[\vec{d}-\vec{h} \big]\mathcal{C}^{-1} \big[\vec{d}-\vec{h} \big]^{\intercal}+C_0 \\
				   &=\ln\Lambda-\frac{1}{2}\vec{d}\mathcal{C}^{-1}\vec{d}^{\intercal}+C_0, \\
\end{aligned}\label{eq:ll}
\end{equation}
where $\ln\Lambda=\vec{d}\mathcal{C}^{-1}\vec{h}-\frac{1}{2}\vec{h}\mathcal{C}^{-1}\vec{h}$ corresponds to the log-likelihood ratio and $C_0$ corresponds to a constant determined by the determinant of the covariance matrix.

%% Fs
From now on, we exclude the extrinsic parameters $(A_{\ell mn},\phi_{\ell mn})$ from $\theta$.
Please note that all extrinsic parameters occur exclusively in $B^{\mu}$ and not in $\vec{h}_{\mu}$.
This allows us to reformulate the log-likelihood ratio as
\begin{equation}
\ln\Lambda(\theta,B^{\mu})=B^{\mu}s_{\mu}(\theta)-\frac{1}{2}B^{\mu}M_{\mu\nu}(\theta)B^{\nu},
\label{eq:llr}
\end{equation}
where we have used the definitions from Eq.~(\ref{eq:rin_split}) and the following conventions: $s_{\mu}=\vec{d}\mathcal{C}^{-1}\vec{h}_{\mu}^{\intercal}$ and $M_{\mu\nu}=\vec{h}_{\mu}\mathcal{C}^{-1}\vec{h}_{\nu}^{\intercal}$.
We then maximize the log-likelihood ratio over parameters $B^{\mu}$ by solving
\begin{equation}
\frac{\partial \ln\Lambda(\theta, B^{\lambda})}{\partial B^{\nu}}=s_{\nu}-B^{\mu}M_{\mu\nu}=0.
\end{equation}
Straightforwardly, we find 
\begin{equation}\label{eq:fsB}
B^{\mu}=(M^{-1})^{\mu\nu}s_{\nu}. 
\end{equation}
Subsequently, after substituting back into Eq.~(\ref{eq:llr}) and defining the $\mathcal{F}$-statistic as $\mathcal{F} = \ln\Lambda$, we find that the $\mathcal{F}$-statistic can be easily calculated using
\begin{equation}
\mathcal{F}(\theta)=\frac{1}{2}s_{\mu}(M^{-1})^{\mu\nu}s_{\nu}.
\label{eq:Fs}
\end{equation}
The $\theta$ that maximizes $\mathcal{F}$, and hence $\Lambda$, therefore gives the parameters that are taken as the underlying intrinsic parameters of the \ac{GW} source. The assertion that $s_{\mu}$ and $M_{\mu\nu}$ in Eq.~(\ref{eq:Fs}) can be substituted with the summation of $s^1_{\mu}+s^2_{\mu}+...+s^{N_{\rm det}}_{\mu}$ and $M^1_{\mu\nu}+M^2_{\mu\nu}+...+M^{N_{\rm det}}_{\mu\nu}$ respectively, for a scenario encompassing $N_{\rm det}$ distinct detectors, is readily demonstrable \citep{Cutler:2005hc}.

%% advantage
As can be seen in Eq.~(\ref{eq:Fs}), the inclusion of additional modes does not result in an expansion of the parameter space. Typically, $\theta$ embodies seven parameters, namely $(\text{RA}, \text{DEC}, t_\text{c},\psi,\iota,M_f,\chi_f)$; these represent two sky position angles, geocentric reference time, polarization angle, inclination angle, final mass and final spin respectively. In the context of \ac{TD} ringdown analyses, it is customary to fix $(\text{RA}, \text{DEC}, t_\mathrm{c}, \psi,\iota)$ based on other analyses, such as results derived from a comprehensive inspiral-merger-ringdown analysis. Notably in the sky-averaged scenario, there is no requirement to consider $({\rm RA ,DEC}, \psi)$. This implies that if ${t_\text{c}}$ and $\iota$ are fixed then only two parameters are needed for further analysis. 
Therefore, we analytically compute the log-likelihood, which then yields posterior distributions after normalization, assuming uniform priors on the remnant mass and spin.

Note that using a frequentist statistic, like the maximum-likelihood $\fss$, often implicitly assumes some choice of prior in the context of Bayesian marginalisation \citep{searleetal2008, searleetal2009, Prix:2009tq}. For the $\fss$, the implicit priors on the amplitude parameters are uniform, which causes a bias towards larger amplitudes and consequently results in a lower detection probability at fixed false alarm probability. Nevertheless, this effect was found to be small at least in the context of continuous GWs \citep{Prix:2009tq}. Therefore, we proceed with the assumption that it is safe to have uniform priors on the parameters used in the ringdown analysis. Moreover, we show later in Fig.~\ref{fig:ns-analytic} that we are able to recover our injections without problems, justifying the assumption. We would like to further investigate the effects of different choices of priors in future studies.

%--------------------------------------------------------
\begin{figure}
\plotone{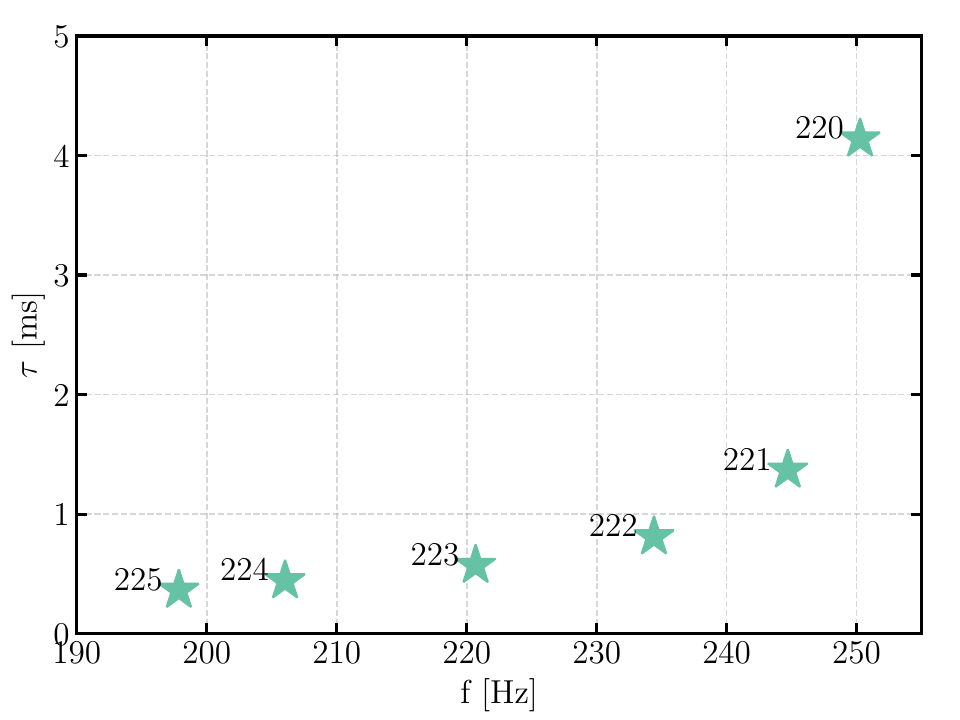}
\caption{
The oscillation frequencies and damping times of the fundamental mode and five overtone modes are presented, corresponding to a final black hole mass of $M_f=68.2\Msun$ and a spin parameter of $\chi_f=0.69$. The numerical values adjacent to the markers denote distinct quasinormal modes, represented in the form $\ell mn$.
}\label{fig:f_taus}
\end{figure}
%--------------------------------------------------------

In Fig.~\ref{fig:f_taus}, we present the oscillation frequencies and damping times of the fundamental mode and various overtone modes for $M_f=68.2\Msun$ and $\chi_f=0.69$. The oscillation frequencies between each pair of adjacent overtones exhibits a close proximity, particularly for the $224$ and $225$ modes. The relative discrepancy in the oscillation frequencies of these two modes is approximately $4\%$. An analogous inference can be drawn for the damping time, where a relative discrepancy of approximately $18\%$ is observed between these two modes.

As depicted in Fig.~\ref{fig:f_taus}, overtone modes exhibit an increased decay rate with increasing order. The early-stage ringdown signal is predominantly governed by these higher overtone modes, which also possess larger amplitudes \citep{Overtone_PRX_Giesler2019}. Consequently, a more comprehensive inclusion of overtone modes becomes necessary when matching data from earlier times. In such instances, we ascertain that up to five overtones should be incorporated into the ringdown waveform if matched with data commencing at $\Delta t = 3$M post-peak, where M$=68.2\Msun$ \footnote{Using the geometric units ($G=c=1$), the characteristic timescale associated with $3\times 68.2\Msun$ is approximately $1\,\rm ms$.}. 
Conversely, optimal matching for a ringdown waveform comprising solely of the fundamental mode should commence at $28$M following peak amplitude. For each additional overtone mode considered, an extra $5$M worth of data is included in our analysis. It is important to note that minor alterations in the start time for each case do not significantly impact our primary conclusions, as shown in discussions related to Fig.~\ref{fig:mfsf2}.

%% ------------------------
\section{Implementation of the $\mathcal{F}$-statistic}\label{sec:results}
Utilizing Eq.~(\ref{eq:ll}) and Eq.~(\ref{eq:Fs}), Bayesian inferences are conducted employing both the TTD method and the $\mathcal{F}$-statistic. In each instance, we fix the reference time $t_\text{c}$ and inclination angle $\iota$, congruent with the injection. Assumptions of flat priors for the other parameters are made within these ranges: $M_f\in[50,90]\Msun$, $\chi_f\in[0.4,0.9]$, $A_{22n}\in[0,250]\times 10^{-20}$, and $\phi_{22n}\in[0,2\pi)$. The data under simulation span a duration of $2048~\text{s}$ at a sample rate of $2048~\text{Hz}$. 

\subsection{Comparison between the analytical solution and the nested sampling solution}\label{subsec:com1}
For the $\fss$, there are two methods to obtain posterior distributions. 
The first is an analytical approach where the log-likelihood is computed. 
We uniformly partition the mass range $[50,90]\Msun$ and the spin range $[0.4,0.9]$ into a $100 \times 100$ grid, calculating the log-likelihood at each grid point. 
This computation utilizes the {\sc Multiprocessing} package \citep{Hunt2019} with $10$ threads.
\footnote{The {\sc python} version is 3.9.} 
The second method employs the nested sampling algorithm implemented in the {\sc Bilby} package \citep[v2.1.1;][]{Ashton_APJ2019}, also used in the TTD method. 
In both methods, Bayesian inferences are conducted using the {\sc dynesty} sampler \citep[v2.1.2;][]{Dynesty_MNRAS_Speagle2020}, with $1000$ live points and a maximum of $1000$ Markov chain steps, setting the {\it queuesize} parameter to $10$.

%% ------------------------
\begin{figure*}
\plottwo{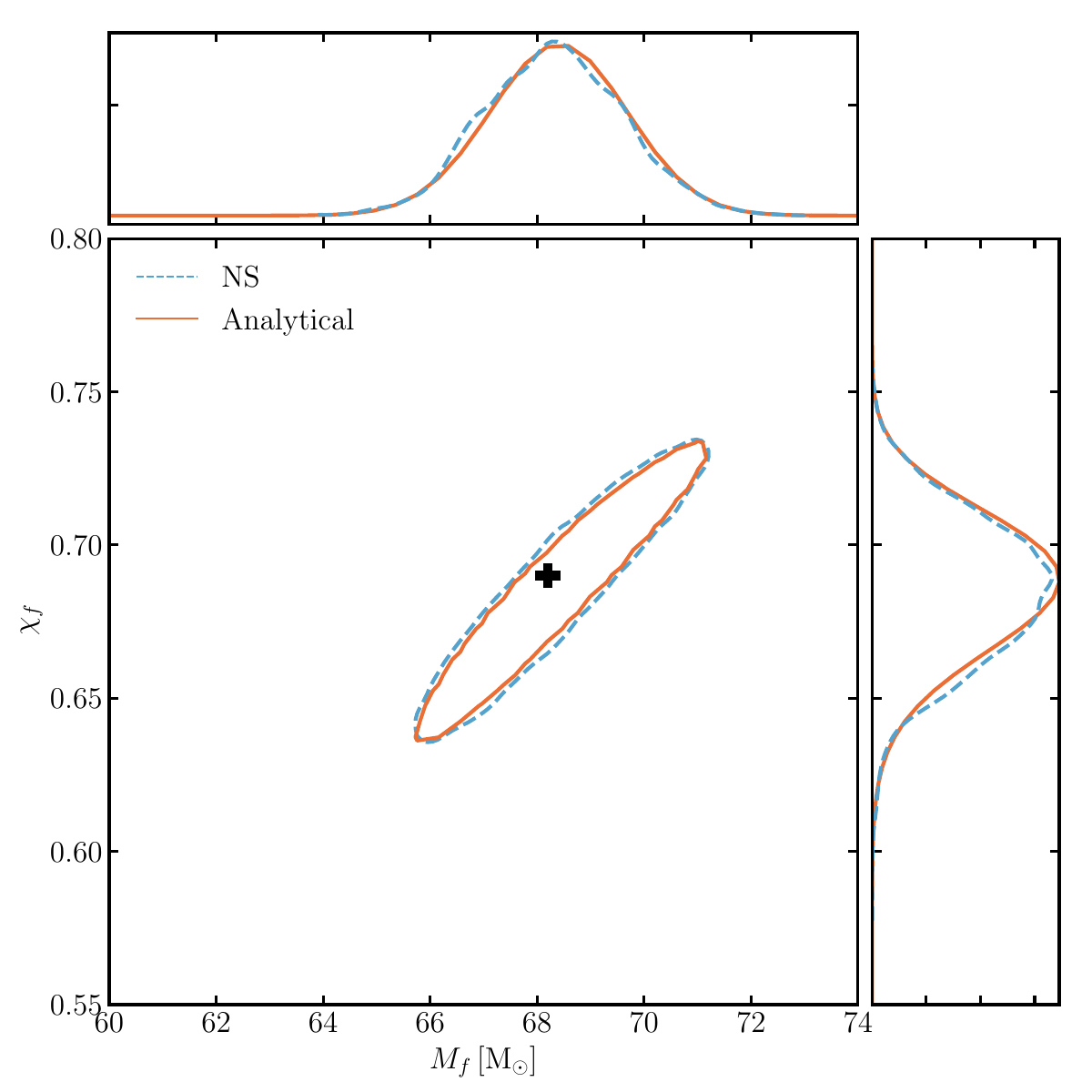}{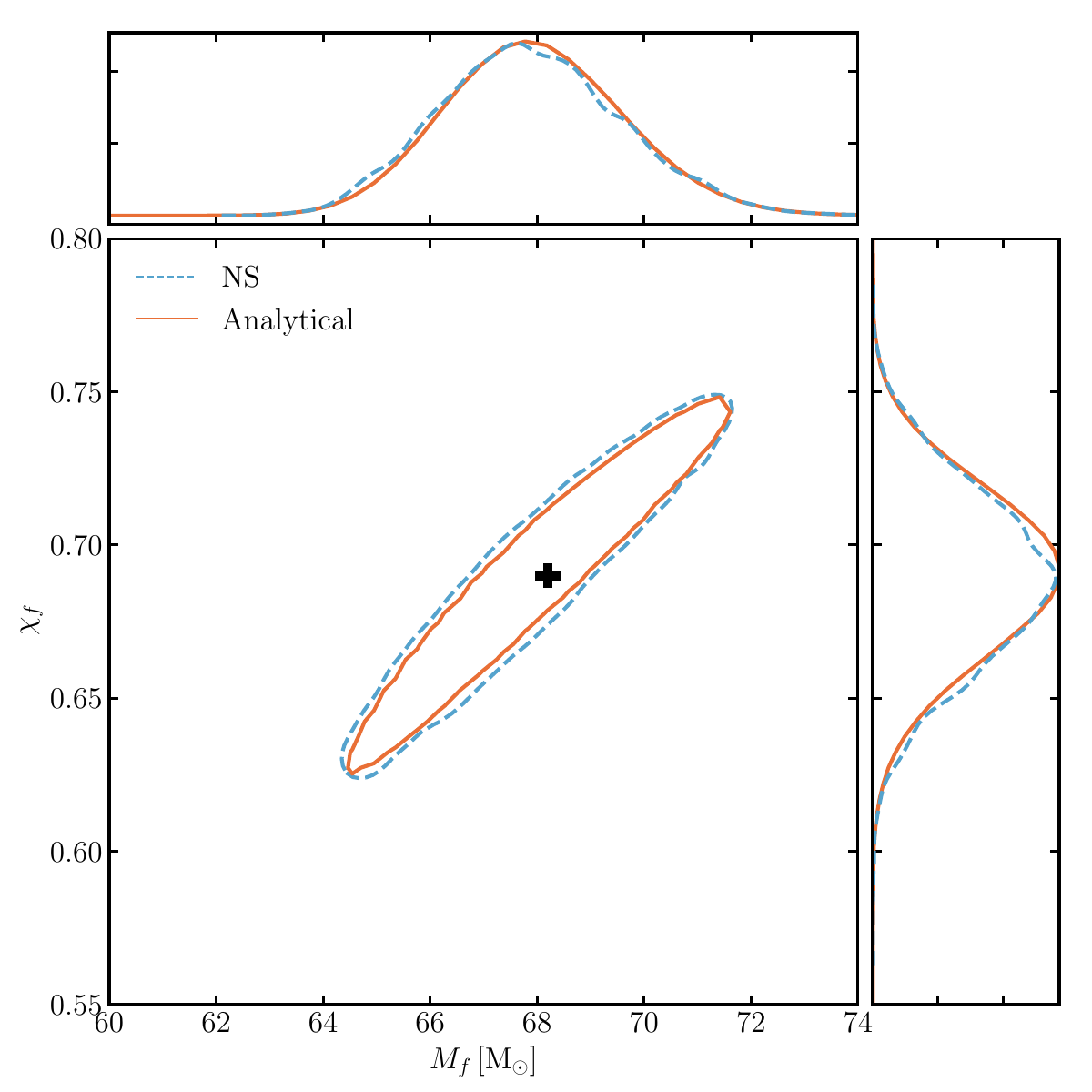}
\caption{
The posterior distributions of the redshifted final mass $M_f$ and final spin $\chi_f$, as determined by the nested sampling solution (dashed blue curves) and the analytical solution (solid red curves) of the $\mathcal{F}$-statistic, are presented. 
Results in the left (right) panel relate to the case with $N=0$ ($N=6$), assuming the ringdown signal starts from $\Delta t = 28$M ($\Delta t = 3$M) after the peak amplitude, where M$=68.2\Msun$ denotes the remnant mass of the injected signal.
The contours illustrate the $90\%$-credible regions for the remnant's parameters, while one-dimensional ($1$D) posteriors for $M_f$ and $\chi_f$ are displayed in the top and right-hand panels respectively. 
The black ``$+$" marker represents the injected values for the redshifted final mass and final spin.
}\label{fig:ns-analytic}
\end{figure*}
%% ------------------------

Firstly, we assess the consistency between the analytical solution and the nested sampling solution for the $\fss$ method. 
We perform ringdown analyses separately using these two solutions for each case. 
As shown in Fig.~\ref{fig:ns-analytic}, we present results for cases with different overtone numbers, $N=1$ and $N=6$, and start times, $\Delta t=28$M and $\Delta t=3$M. 
These solutions show consistent results across different cases. 
Comparisons of other cases are not shown, as they yield consistent results aligned with intuition and theoretical expectations.
Each solution has its advantages and disadvantages. 
The analytical solution can rapidly produce the joint posterior \ac{PDF} within seconds; however, it cannot directly derive the relative \ac{PDF} of the log-likelihood. 
In other words, the analytical solution provides analytical \ac{PDF}s of the parameters rather than parameter samples. 
However, we need samples and the corresponding log-likelihood to obtain the \ac{PDF} of the log-likelihood for comparison with the TTD method, as shown in Fig.~\ref{fig:lls}. 
In contrast, the nested sampling solution, though slower (usually taking hours), directly yields the \ac{PDF} of the log-likelihood.
Henceforth, we will not distinguish results from these two solutions, as they are consistent with each other.

%% ------------------------
\begin{figure*}
\plottwo{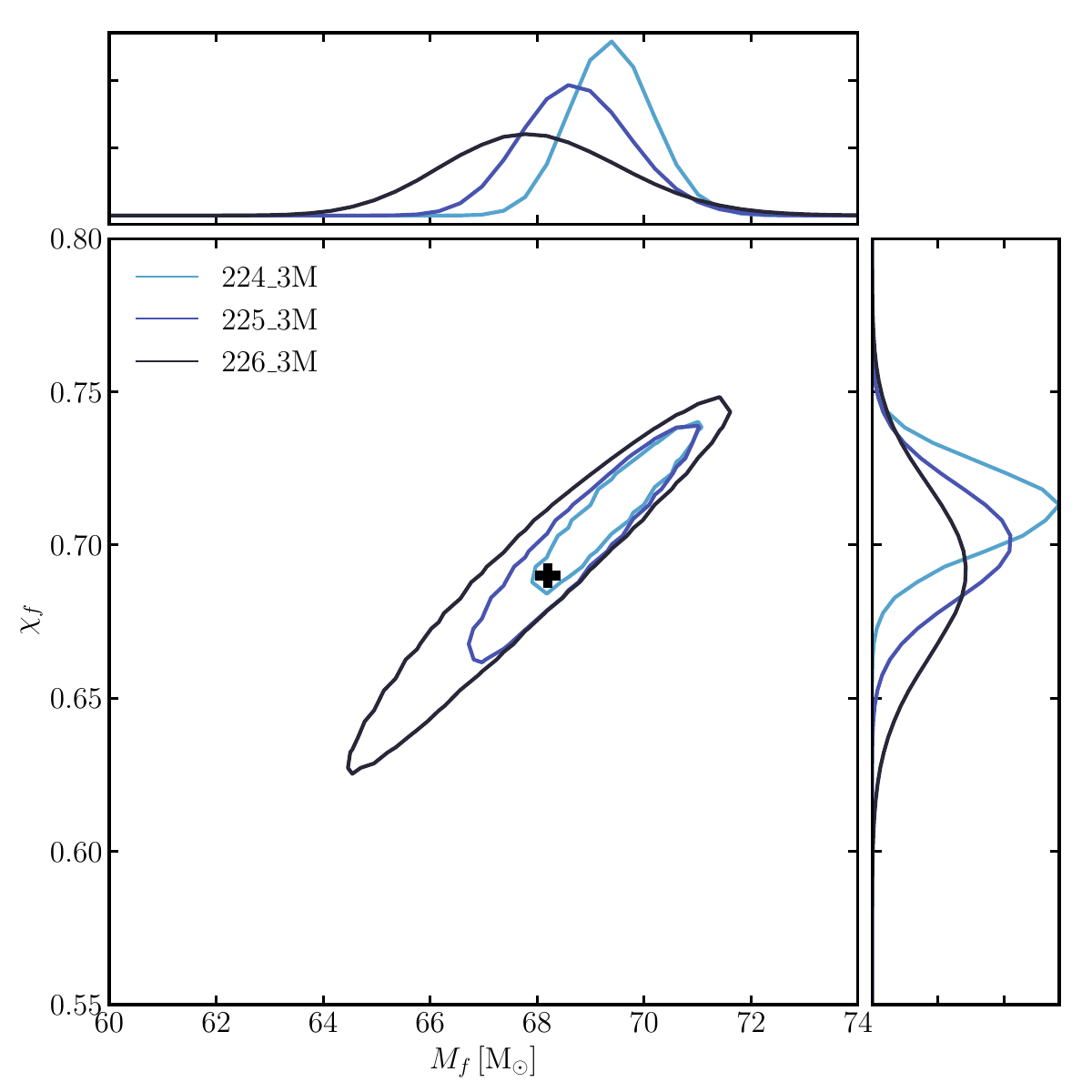}{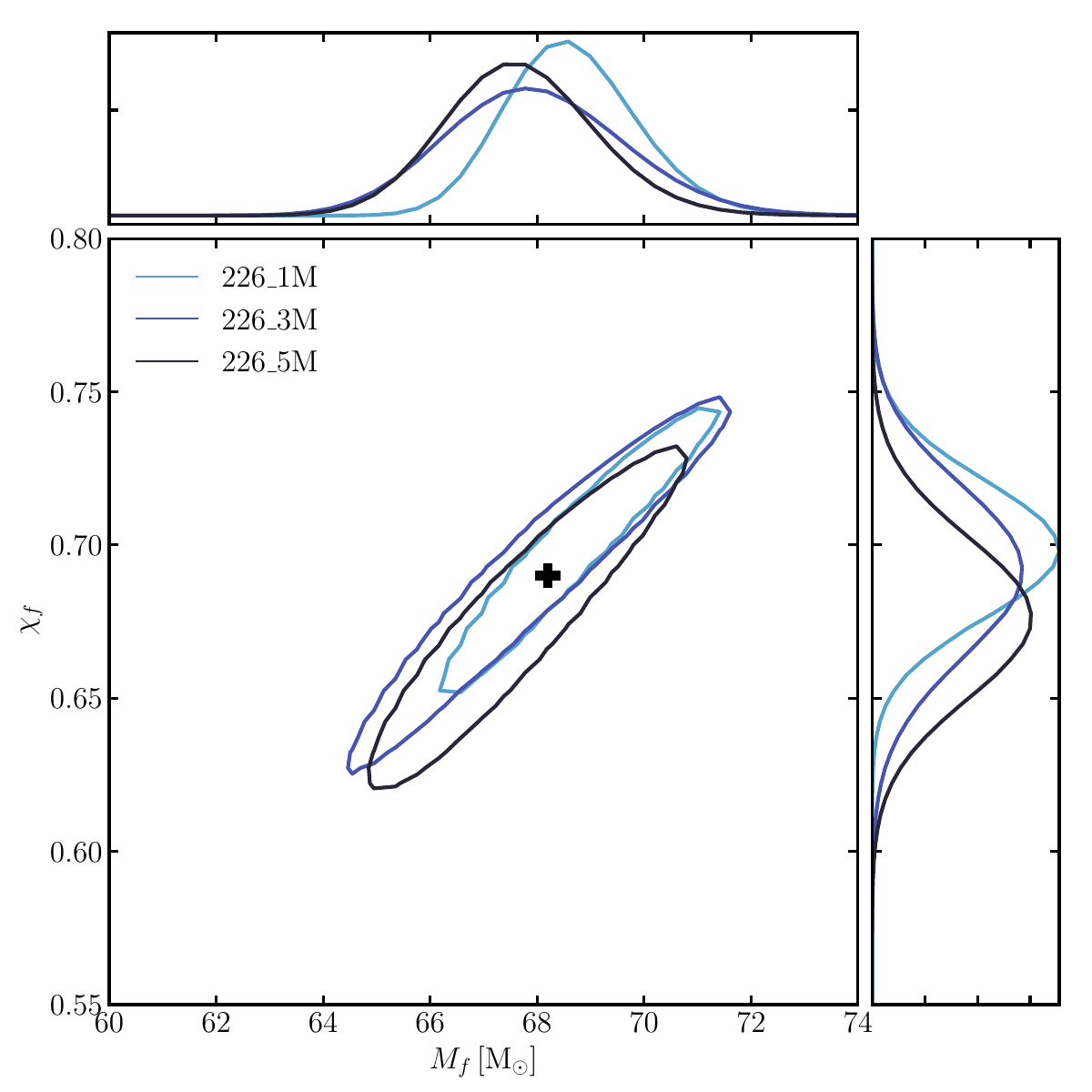}
\caption{
The posterior distributions of the redshifted final mass $M_f$ and final spin $\chi_f$, as determined by different numbers of overtone modes (left panel) and different starting times (right panel). 
The contours illustrate the $90\%$-credible regions for the remnant's parameters, while one-dimensional ($1$D) posteriors for $M_f$ and $\chi_f$ are displayed in the top and right-hand panels respectively. 
We consider varying numbers of overtone modes, initiated at different post-peak times, denoted by $\ell mN\_\Delta t$.
For instance, a label of $226\_3$M signifies that a waveform incorporating the fundamental mode and five overtone modes is used in the simulated strain data, commencing at $\Delta t = 3$M post-peak.
The black ``$+$" marker represents the injected values for the redshifted final mass and final spin.
These results are based on the $\fss$ method. 
}\label{fig:mfsf2}
\end{figure*}
%% ------------------------

\subsection{The choice of the number of modes and the start time}\label{subsec:choice}
To bolster the robustness of our conclusions, we have conducted additional analyses for scenarios with different quantities of overtone modes when $\Delta t=3$M, and for those with varying commencement times when $N=6$. As depicted in Fig.~\ref{fig:mfsf2}, constraints for scenarios involving $4$ to $6$ modes exhibit a strong bias when analyses are initiated at $3$M post-peak. Constraints derived from the case labeled as $224\_3$M demonstrate a greater stringency while also gravitating towards an area characterized by increased mass and amplified spin. This is logically consistent given that the ringdown waveform in this particular scenario does not incorporate higher order overtone modes, which typically possess lower frequencies and shorter damping times that can be emulated by a signal featuring elevated mass and enhanced spin.

In the instance of $226\_3$M ($226\_5$M), the final mass and spin are $68.6^{+1.6}_{-2.0}\Msun$ ($67.4^{+2.4}_{-2.0}\Msun$) and $0.70^{+0.03}_{-0.04}$ ($0.67^{+0.05}_{-0.04}$), respectively, with an alignment probability to the true values at $86.2\%$ ($83.0\%$). The alignment probability from the $226\_5$M mode is marginally lower than that in the case of the $226\_3$M mode.
All constraints presented throughout represent a $90\%$ credible level unless otherwise specified.
For the scenario of $226\_1$M, the final mass and spin are found to be $68.6^{+2.0}_{-2.0}\Msun$ and $0.70^{+0.04}_{-0.04}$, respectively, accompanied by an alignment probability of $88.2\%$. Despite this case appearing to provide a better match to the injected values, it should be noted that the posterior distribution naturally favors regions characterized by higher masses and larger spins when the start time is earlier, since higher order overtones have lower $\tau$. Caution must therefore be exercised when incorporating additional data without further supporting information indicating its validity. 

Consequently, we adopt an informed approach in our analyses; we assume that the ringdown waveform with $N=6$ commences at $3$M after the peak amplitude.
Overall, the start times chosen in our main text can be deemed reasonable given the minor discrepancies observed upon slight shifts in commencement time.

%--------------------------------------------------------
\begin{figure}
\plotone{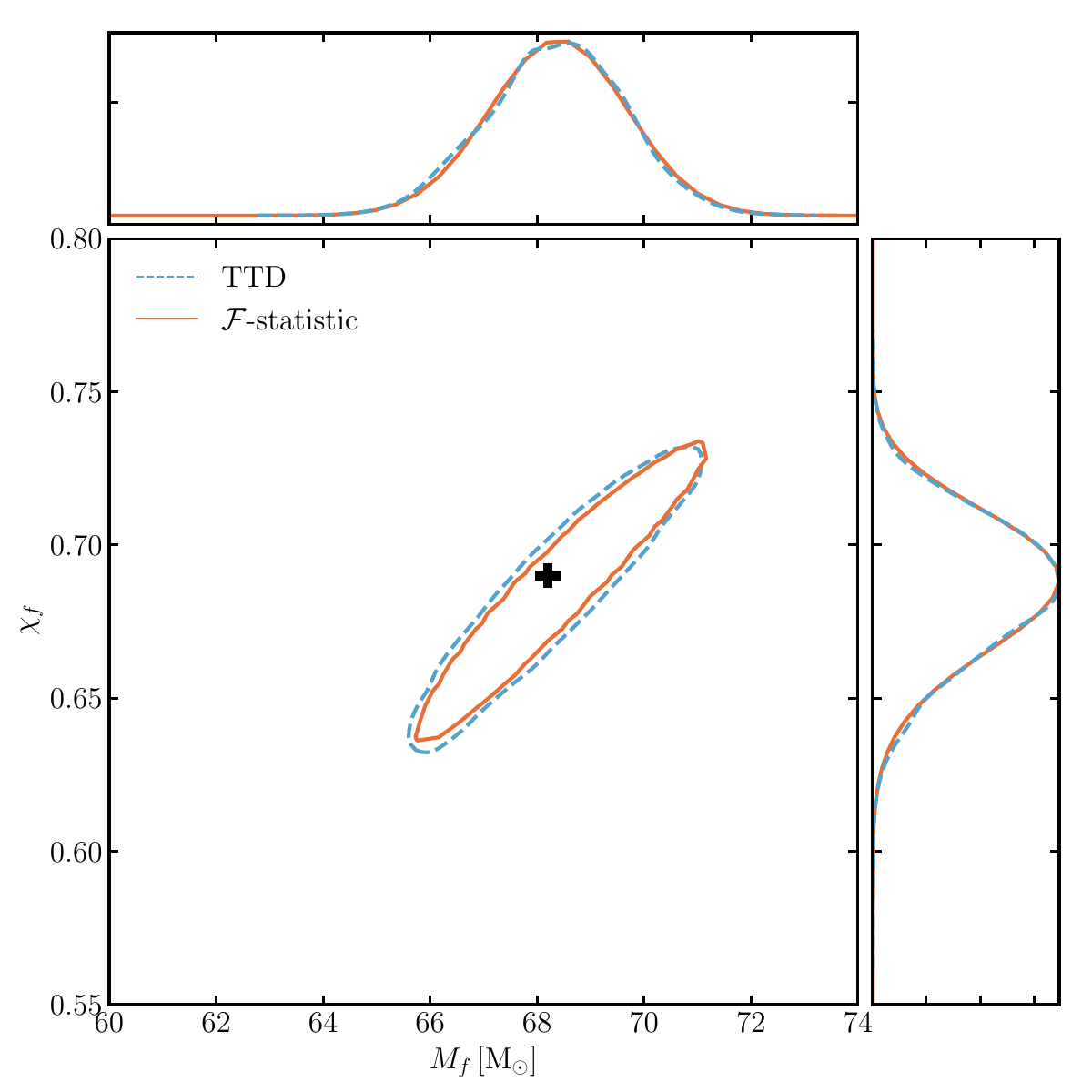}
\caption{
The posterior distributions of the redshifted final mass $M_f$ and final spin $\chi_f$, similar to Fig.~\ref{fig:mfsf1}.
We compare results of the case $221\_28$M for both the TTD method and the $\fss$ method, which are labeled by ``TTD" and ``$\fss$", respectively.
}\label{fig:com221}
\end{figure}
%--------------------------------------------------------

\subsection{Comparison between the TTD method and the $\fss$ method}\label{subsec:com2}
In the instance of $221\_28$M, denoting that parameter estimation commences $\Delta t = 28$M after the peak and is solely governed by the fundamental mode, outcomes derived from both the TTD method and the $\fss$ method exhibit consistency. As depicted in Fig.~\ref{fig:com221}, constraints on the final mass and spin are determined to be $68.4^{+2.1}_{-2.2}\Msun$ ($68.2^{+2.4}_{-2.4}\Msun$) and $0.69^{+0.04}_{-0.04}$ ($0.68^{+0.04}_{-0.04}$) respectively for the TTD method ($\fss$ method). The marginal discrepancy between these results can be attributed to a reduced parameter space when using the $\fss$ method.

%% ------------------------
\begin{figure*}
\plottwo{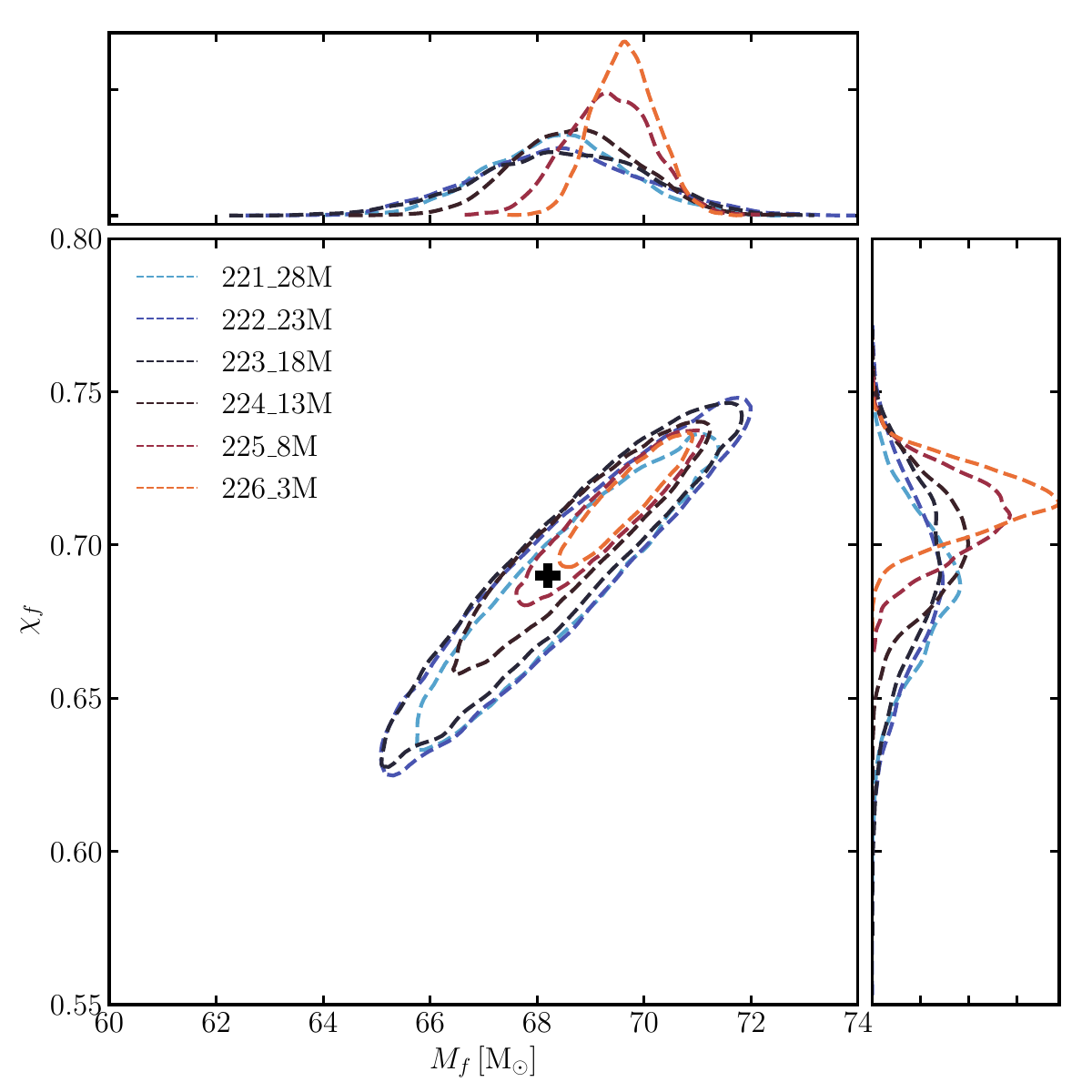}{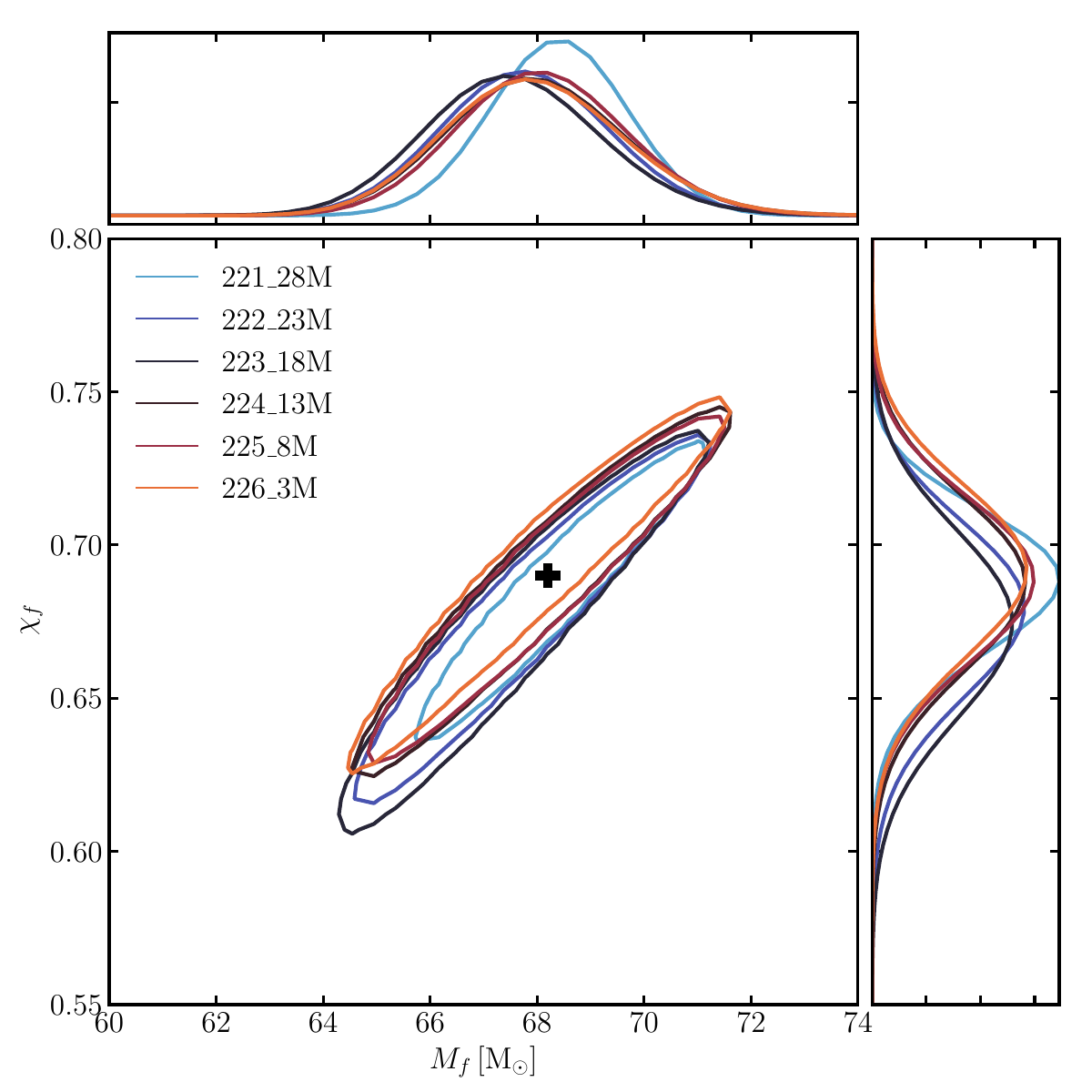}
\caption{
The posterior distributions of the redshifted final mass $M_f$ and final spin $\chi_f$, as determined by the TTD method (left panel) and the $\mathcal{F}$-statistic (right panel), are presented. 
The contours illustrate the $90\%$-credible regions for the remnant's parameters, while one-dimensional ($1$D) posteriors for $M_f$ and $\chi_f$ are displayed in the top and right-hand panels respectively. 
The black ``$+$" marker represents the injected values for the redshifted final mass and final spin.
Labels are similar with those in Fig.~\ref{fig:mfsf2}.
}\label{fig:mfsf1}
\end{figure*}
%% ------------------------

As we show above, the outcomes of both the TTD method and the $\mathcal{F}$-statistic are in agreement when solely considering the fundamental mode.
However, comparing constraints from the TTD method with those derived from the $\mathcal{F}$-statistic becomes challenging when more than three overtones are incorporated into the ringdown waveform analysis, as depicted in Fig.~\ref{fig:mfsf1}. 
For instance, when four overtones are included in such an analysis, constraints from the TTD method yield ($69.3^{+1.3}_{-1.3}\Msun$, $0.71^{+0.02}_{-0.02}$) for ($M_f$, $\chi_f$), with a probability of alignment with the true values ($68.2\Msun$, $0.69$) being $31.9\%$. 
In contrast, using the $\mathcal{F}$-statistic results in constraints of ($67.8^{+2.8}_{-2.4}\Msun$, $0.69^{+0.04}_{-0.05}$), offering an increased probability of concurrence with the true values of approximately $96.6\%$. 

The discrepancy is notably amplified when five overtones are incorporated into the ringdown waveforms. The constraints on the final mass and spin exhibit significant discrepancies between the two methods, with values calculated as $69.6^{+1.0}_{-0.9}\Msun$ ($67.8^{+2.8}_{-2.8}\Msun$) and $0.71^{+0.02}_{-0.02}$ ($0.69^{+0.05}_{-0.05}$) for the TTD method ($\mathcal{F}$-statistic method). 
The probability of alignment with the true values is drastically reduced (improved) to $3.6\%$ ($85.6\%$) for the TTD method ($\mathcal{F}$-statistic method).

For the other cases illustrated in Fig.~\ref{fig:mfsf1}, there is general agreement with the true values exceeding $70\%$. Notably, constraints derived from the $\mathcal{F}$-statistics become less influenced as more overtone modes are incorporated into the waveform models. This suggests that the additional data contributions are negligible compared to the introduction of extra parameters due to the inclusion of overtone modes. This aligns with our expectations, as higher overtone modes decay faster, contributing less to the \ac{SNR}.
In some prior studies \citep{Overtone_PRX_Giesler2019,Isi:2019aib}, researchers concluded that including higher overtone modes could significantly improve constraints on remnants. However, this conclusion is likely biased due to a ``bug"—a poor choice of the re-sampling algorithm—in their noise estimation method.\footnote{The power spectral density exhibits an apparent decline near the Nyquist frequency due to this poor choice, which is unphysical and results in biased estimations in time-domain Bayesian inference. This issue can be mitigated by employing a Butterworth filter during the re-sampling process, as shown in \citet{Wang:2023mst} and \citet{Wang:2024liy}.} Our tests indicate that their method struggles to pass the consistency check between time-domain and frequency-domain Bayesian inferences unless this ``bug" is addressed \citep{Wang:2024liy}. 
Furthermore, this ``bug" is the primary reason for inconsistencies between the results in \citet{Isi:2019aib} and \citet{Carullo:2023gtf}. 
After addressing this issue, \citet{Wang:2023mst} obtained consistent results across different sampling rates and observed that the improvement is limited when including the first overtone mode.

%% ------------------------
\begin{figure*}
\gridline{\fig{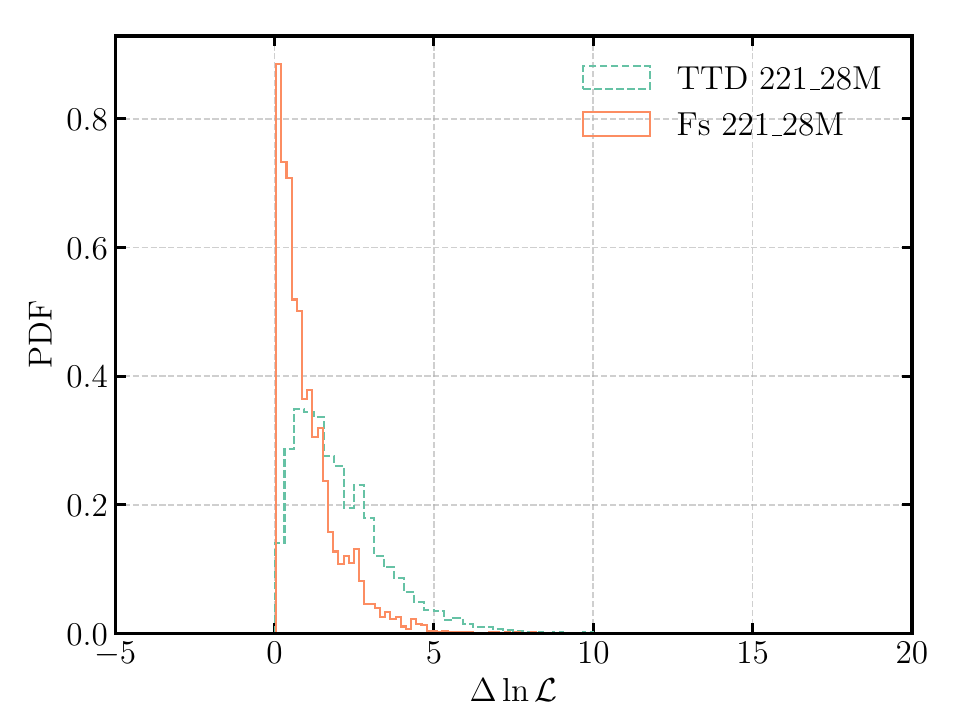}{0.48\textwidth}{(a)}
          \fig{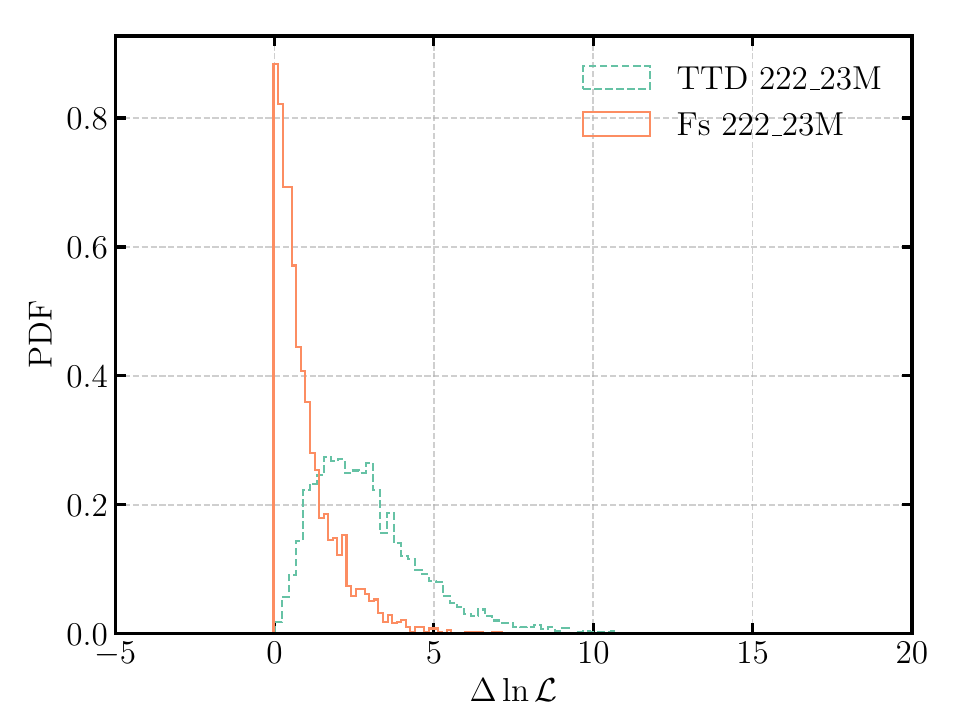}{0.48\textwidth}{(b)}
          }
\gridline{\fig{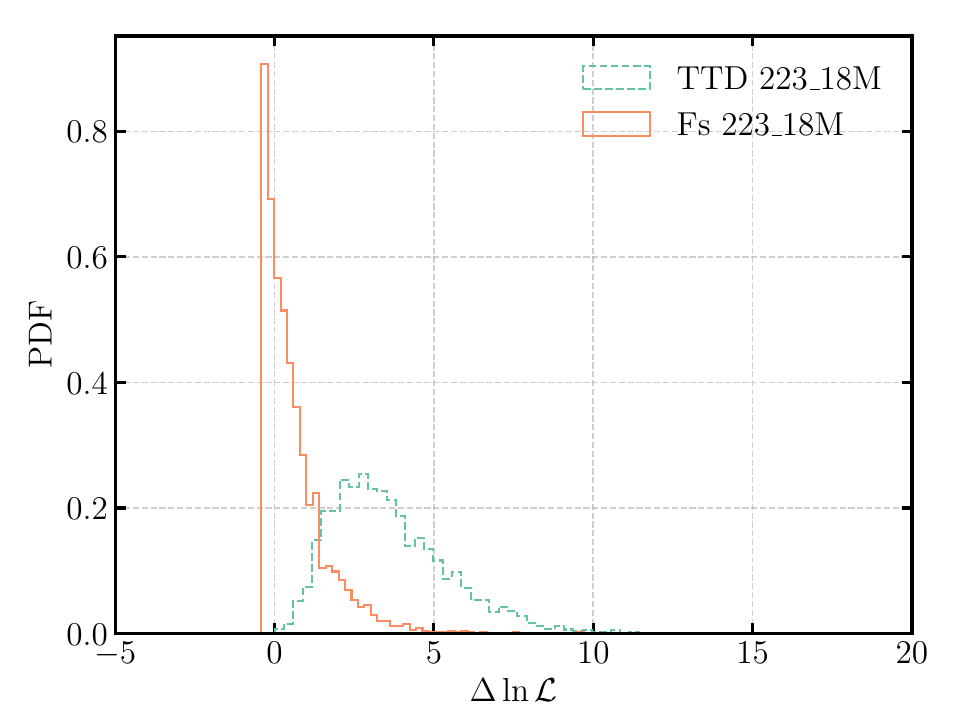}{0.48\textwidth}{(c)}
          \fig{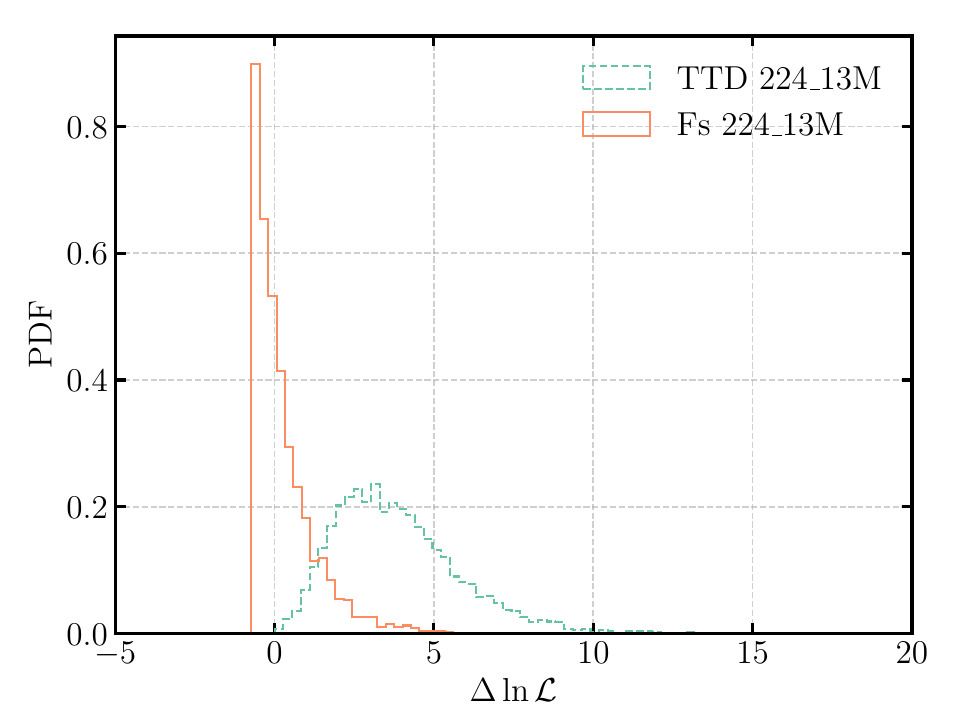}{0.48\textwidth}{(d)}
          }
\gridline{\fig{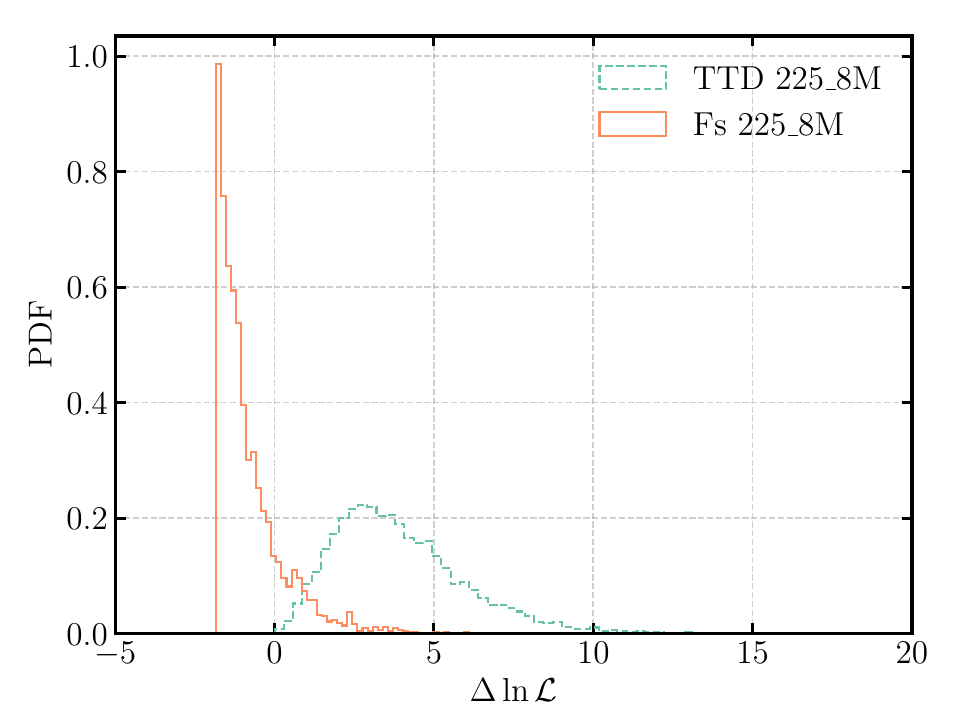}{0.48\textwidth}{(e)}
          \fig{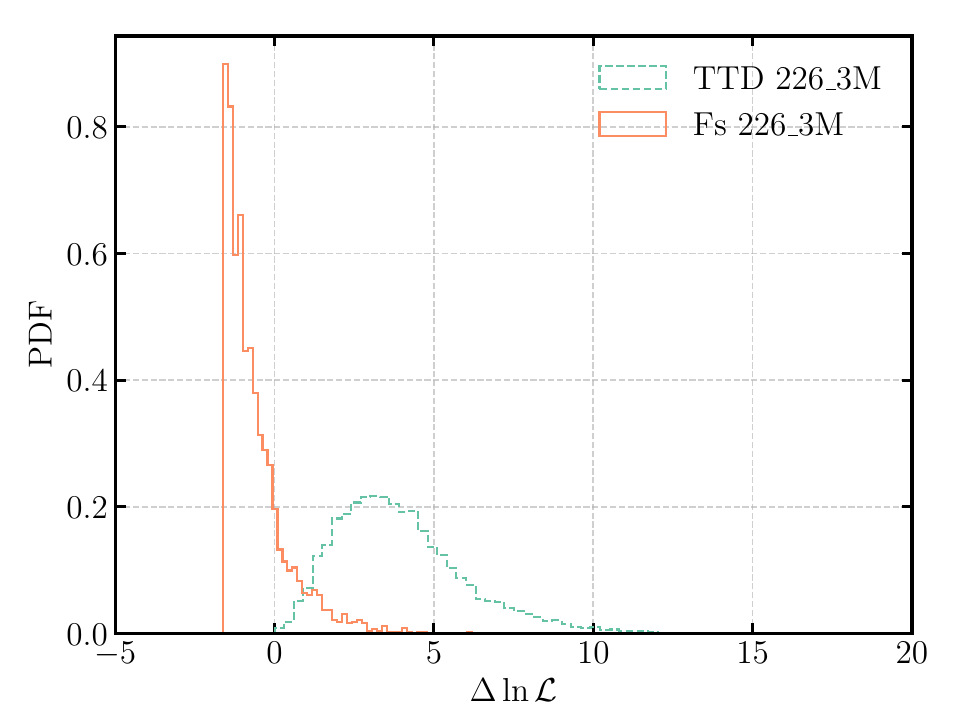}{0.48\textwidth}{(f)}
		  }
\caption{
	Distributions of $\Delta\ln \mathcal{L}=\max(\ln \mathcal{L}_{\rm TTD})-\ln \mathcal{L}$ showing log-likelihood differences between the $\fss$ method (labels starting with `Fs') and the TTD method. Each subfigure corresponds to an individual case that can be found in Fig.~\ref{fig:mfsf1}.
}\label{fig:lls}
\end{figure*}
%% ------------------------

%% \subsection{The comparison of the log-likelihood}
%% Therefore, results in Fig.~\ref{fig:lls} help us to understand results shown in Fig.~\ref{fig:mfsf1}.

To assess the reliability of the $\fss$ method compared to the TTD method, we compare their log-likelihood distributions, as depicted in Fig.~\ref{fig:lls}. The difference in log-likelihoods is defined as
\begin{equation}
    \Delta\ln \mathcal{L} = \max(\ln \mathcal{L}_{\rm TTD}) - \ln \mathcal{L},
\end{equation}
where $\ln \mathcal{L}$ denotes the log-likelihood of different methods and $\max(\ln \mathcal{L}_{\rm TTD})$ represents the maximum log-likelihood among TTD method samples for each case shown in Fig.~\ref{fig:lls}.

\begin{figure*}
	\gridline{\fig{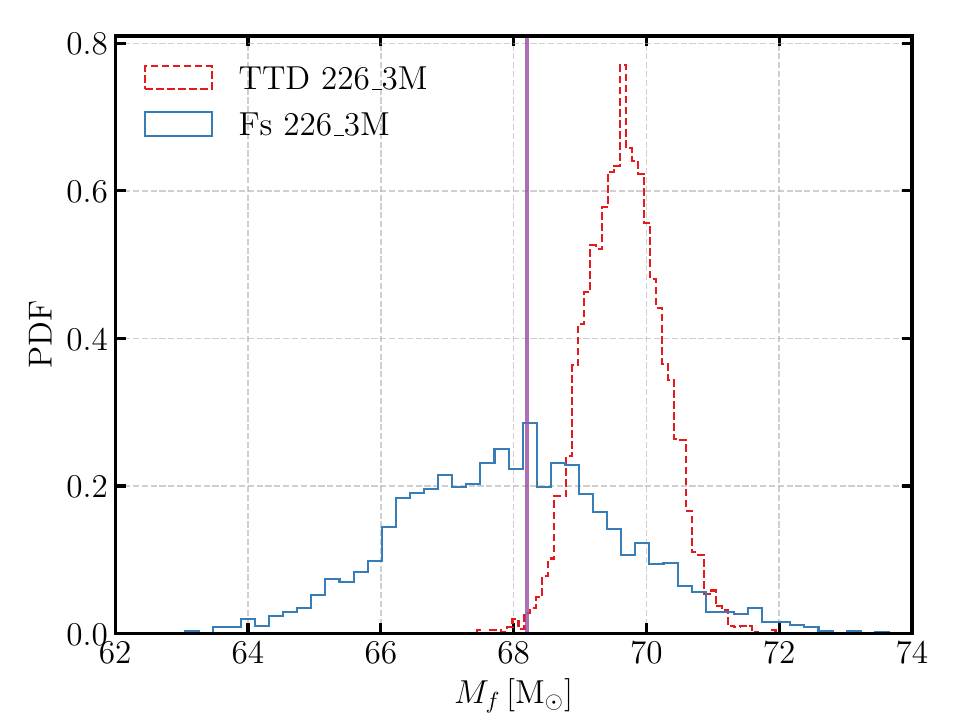}{0.48\textwidth}{(a)}
			  \fig{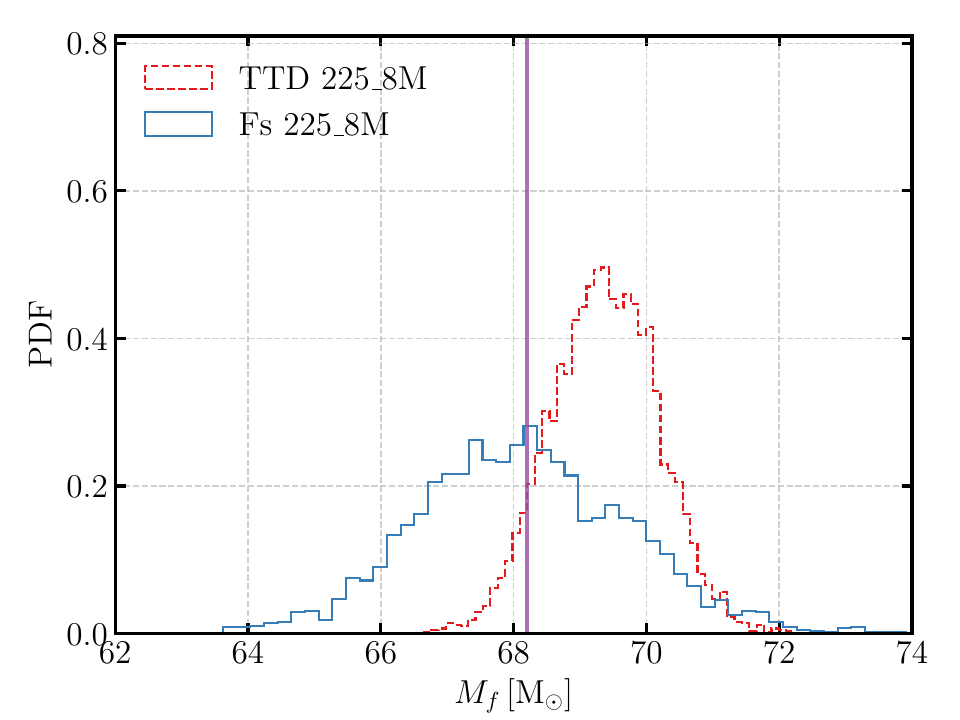}{0.48\textwidth}{(b)}}
	\gridline{\fig{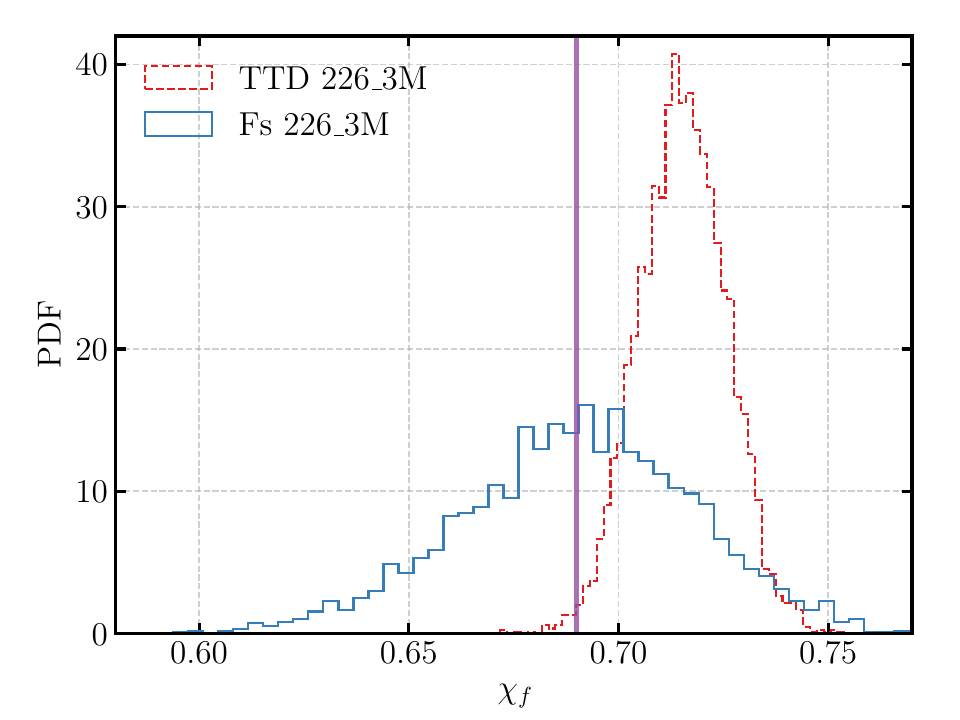}{0.48\textwidth}{(c)}
			  \fig{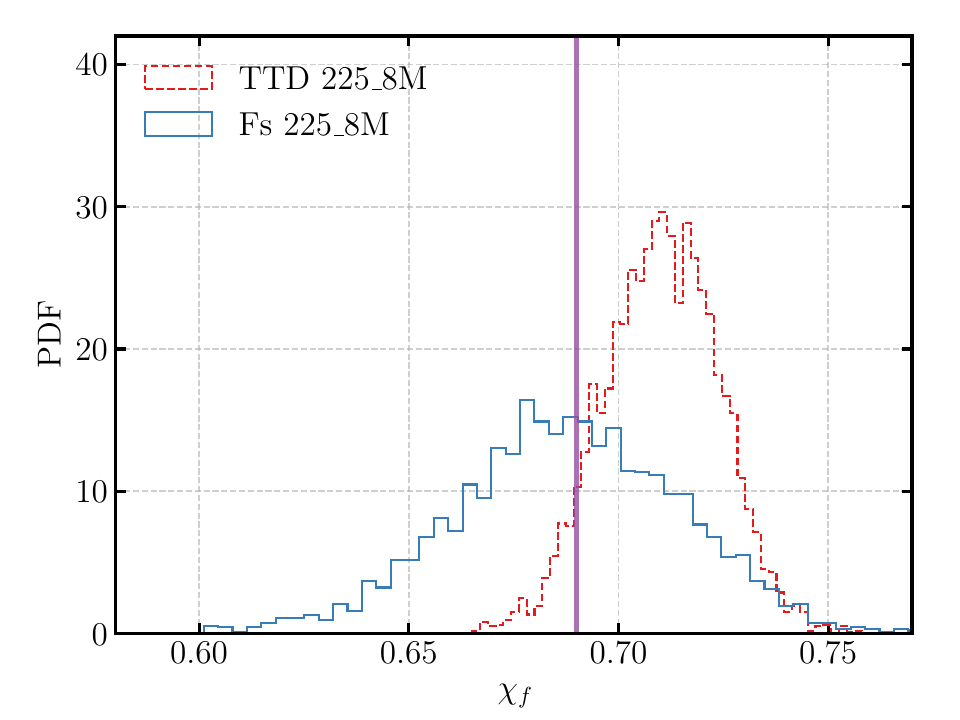}{0.48\textwidth}{(d)}}
\caption{
Histograms of posterior distributions of the redshifted final mass $M_f$ (top two panels) and final spin $\chi_f$ (bottom two panels) from the TTD method (solid blue histograms) and the $\fss$ method (dashed red histograms).
Purple vertical lines indicates the injected values.
Labels of the legend in line with those in Fig.~\ref{fig:lls}.
}\label{fig:fmfs_pdf}
\end{figure*}

We observe that the distributions of $\Delta\ln \mathcal{L}$ for the $221\_28$M case exhibit the closest agreement between the two methods. However, deviations increase when more overtone modes are included. For cases where $N \geq 3$, the $\Delta\ln \mathcal{L}$ distribution of the $\fss$ method tends to peak in the region less than zero, indicating that the maximum log-likelihood provided by the $\fss$ method is higher compared to that of the TTD method in these instances. 
%% Specifically, for $N\geq 5$, the $\Delta\ln \mathcal{L}$ distribution of the TTD method is notably broader than in other cases, while for the $\fss$ method, the distribution for $N\geq 5$ resembles that of other cases. 
For the cases of $N\geq 5$, the $\Delta\ln \mathcal{L}$ distributions peak far from zero showing that the TTD method fails to find the maximum likelihood. Instead, it becomes trapped in local maxima when too many overtone modes are included.
Posterior distributions of redshifted final mass $M_f$ and final spin $\chi_f$ are shown in Fig.~\ref{fig:fmfs_pdf}, generated from both the TTD method and the $\fss$ method, support the conclusion that the $\fss$ method outperforms the TTD method. %% Overall, the TTD method struggles to identify the maximum log-likelihood when a large number of parameters are involved.
This agrees with the results shown in Fig.~\ref{fig:mfsf1}.

%% ------------------------
%% ------------------------

%--------------------------------------------------------
\begin{figure}[ht!]
\plotone{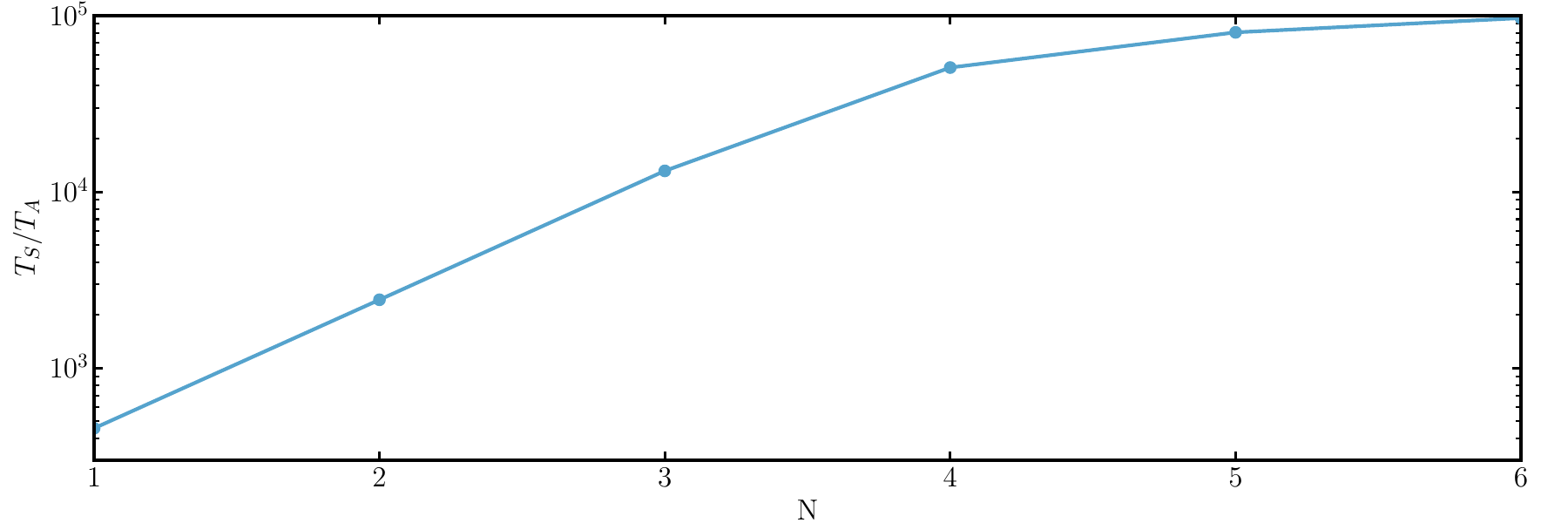}
\caption{
Comparison between the time spent by the TTD method ($T_S$) and the $\mathcal{F}$-statistic method ($T_A$), contingent upon the number of QNMs ($N$) incorporated within the ringdown waveform.
%Constraints of these analyses are shown in Fig.~\ref{fig:mfsf1}.
}\label{fig:ts}
\end{figure}
%--------------------------------------------------------

In Fig.~\ref{fig:ts}, we show the speed-up of the ${\cal F}$-statistic method, demonstrating its superior performance relative to the TTD method for different numbers of modes. For instance, with only the fundamental mode, the $\mathcal{F}$-statistic method operates approximately $500$ times faster than its traditional counterpart. 
Remarkably, it functions nearly $8\times10^4$ times quicker for cases where the ringdown waveform incorporates four or five overtones. This can be attributed to the fact that $\mathcal{F}$-statistic constraints are computed analytically since the parameter space remains the same irrespective of how many overtone modes are included in the waveform model. Conversely, for the TTD method,  each additional overtone introduces two extra parameters. However, a saturation in the speed-up of the $\mathcal{F}$-statistic method is observed when $N\geq 5$, due to an increase in the time needed for the matrix inversion in Eq.~(\ref{eq:Fs}), introduced by additional overtone modes.

%% ------------------------
\section{Discussion and prospects}\label{sec:dp}
In this study, we introduced a novel approach for distinguishing multiple modes with similar frequencies or damping times by constructing the $\mathcal{F}$-statistic for ringdown analyses. We further developed a framework predicated on this solution to facilitate \ac{PE} of ringdown signals in \ac{GW} data. The efficacy of our method was evaluated through PEs performed on an injection test, where a GW150914-like \ac{NR} strain was injected into noise data from \ac{ET}. For comparative purposes, analogous analyses were conducted using the TTD method. Our findings indicate that the TTD method struggles to differentiate contributions from distinct overtone modes particularly when $N\geq 5$. Consequently, results derived from PE exhibit significant bias when five overtone modes are incorporated into the ringdown waveform. Quantitatively, congruence with the injected signal occurs merely at a probability of $3.6\%$. In contrast, the application of the $\mathcal{F}$-statistic not only enhances this probability to $85.6\%$, but also expedites estimation time by about five orders of magnitude.

Our framework presents several distinct advantages. Primarily, within the realm of \ac{GW} data analysis, it addresses the issue of distinguishing between oscillation frequencies and damping times of higher overtone modes that are remarkably close. Furthermore, irrespective of how many modes are incorporated into the ringdown waveform, our parameter space remains the same without any loss of \ac{GW} data information. The framework also retains the benefits associated with TTD methods; its flexibility allows for easy extension to other studies such as testing the no-hair theorem \citep{Isi:2019aib,2021PhRvD.103b4041B}, examining \ac{GR} \citep{2021PhRvD.103l2002A,2021arXiv211206861T,2021PhRvD.104j4063W,Cheung:2020dxo,Mishra:2021waw}, and scrutinizing the \ac{BH} area law \citep{Isi:2020tac}. Lastly, this approach considerably reduces computational costs due to a smaller parameter space.

In other words, the framework presented herein enhances the field of ringdown analysis.
On the one hand, it can be employed in \ac{BH} spectroscopy for real \ac{GW} data detected by the \ac{LVK} Collaboration---analyses utilizing this new framework on the ringdown signal of GW150914 are currently underway.
On the other hand, it is applicable to data analyses based on future detectors such as \ac{ET} \citep{2010CQGra..27s4002P}, Cosmic Explorer \citep{2019BAAS...51g..35R}, Laser Interferometer Space Antenna \citep{LISA_arxiv2017}, TianQin \citep{TQ_2015,TianQin:2020hid}, and Taiji \citep{1093nsrnwx116}.
Furthermore, studies in \citet{Keppel:2012ye} and here motivates us to update the $\mathcal{F}$-statistic for the fully inspiral-merger-ringdown analysis.

%% ------------------------
\begin{acknowledgments}
We thank Yi-Ming Hu for fruitful discussions and the anonymous referee for constructive comments.
This work was supported by the China Postdoctoral Science Foundation (2022TQ0011), the National Natural Science Foundation of China (12247152, 12247180, 11975027, 11991053), the Beijing Natural Science Foundation (1242018), the National SKA Program of China (2020SKA0120300), the Max Planck Partner Group Program funded by the Max Planck Society, the Fundamental Research Funds for the Central Universities and the High-performance Computing Platform of Peking University.
HTW is supported by the Opening Foundation of TianQin Research Center. 
\end{acknowledgments}

\bibliographystyle{aasjournal}
\bibliography{fs0305}

\begin{thebibliography}{}
\expandafter\ifx\csname natexlab\endcsname\relax\def\natexlab#1{#1}\fi
\providecommand{\url}[1]{\href{#1}{#1}}
\providecommand{\dodoi}[1]{doi:~\href{http://doi.org/#1}{\nolinkurl{#1}}}
\providecommand{\doeprint}[1]{\href{http://ascl.net/#1}{\nolinkurl{http://ascl.net/#1}}}
\providecommand{\doarXiv}[1]{\href{https://arxiv.org/abs/#1}{\nolinkurl{https://arxiv.org/abs/#1}}}

\bibitem[{Abadie {et~al.}(2010)}]{LIGOScientific:2010tql}
Abadie, J., {et~al.} 2010, Astrophys. J., 722, 1504,
  \dodoi{10.1088/0004-637X/722/2/1504}

\bibitem[{Abbott {et~al.}(2019{\natexlab{a}})}]{LIGO_PRX2019}
Abbott, B.~P., {et~al.} 2019{\natexlab{a}}, Phys. Rev. X, 9, 031040,
  \dodoi{10.1103/PhysRevX.9.031040}

\bibitem[{Abbott {et~al.}(2019{\natexlab{b}})}]{LIGOScientific:2019yhl}
---. 2019{\natexlab{b}}, Phys. Rev. D, 100, 024004,
  \dodoi{10.1103/PhysRevD.100.024004}

\bibitem[{{Abbott} {et~al.}(2020)}]{2020PhRvL.125j1102A}
{Abbott}, R., {et~al.} 2020, \prl, 125, 101102,
  \dodoi{10.1103/PhysRevLett.125.101102}

\bibitem[{{Abbott} {et~al.}(2021)}]{LIGO_O3a_PRX2020}
---. 2021, Phys. Rev. X, 11, 021053, \dodoi{10.1103/PhysRevX.11.021053}

\bibitem[{Abbott {et~al.}(2021{\natexlab{a}})}]{2021arXiv211206861T}
Abbott, R., {et~al.} 2021{\natexlab{a}}, arXiv e-prints, arXiv:2112.06861.
\newblock \doarXiv{2112.06861}

\bibitem[{Abbott {et~al.}(2021{\natexlab{b}})}]{2021PhRvD.103l2002A}
---. 2021{\natexlab{b}}, \prd, 103, 122002, \dodoi{10.1103/PhysRevD.103.122002}

\bibitem[{Abbott {et~al.}(2021{\natexlab{c}})}]{LIGOScientific:2021mwx}
---. 2021{\natexlab{c}}, Astrophys. J., 921, 80,
  \dodoi{10.3847/1538-4357/ac17ea}

\bibitem[{Abbott {et~al.}(2023)}]{KAGRA:2021vkt}
---. 2023, Phys. Rev. X, 13, 041039, \dodoi{10.1103/PhysRevX.13.041039}

\bibitem[{Abedi {et~al.}(2023)Abedi, Capano, Kastha, Nitz, Wang, Westerweck,
  Nielsen, \& Krishnan}]{Abedi:2023kot}
Abedi, J., Capano, C.~D., Kastha, S., {et~al.} 2023, Phys. Rev. D, 108, 104009,
  \dodoi{10.1103/PhysRevD.108.104009}

\bibitem[{{Amaro-Seoane} {et~al.}(2017){Amaro-Seoane}, {Audley}, {Babak},
  {Baker}, {et~al.}}]{LISA_arxiv2017}
{Amaro-Seoane}, P., {Audley}, H., {Babak}, S., {Baker}, J., {et~al.} 2017,
  ArXiv e-prints, arXiv:1702.00786.
\newblock \doarXiv{1702.00786}

\bibitem[{Ashton {et~al.}(2019)}]{Ashton_APJ2019}
Ashton, G., {et~al.} 2019, ApJS, 241, 27, \dodoi{10.3847/1538-4365/ab06fc}

\bibitem[{Babak {et~al.}(2015)Babak, Gair, \& Cole}]{Babak:2014kqa}
Babak, S., Gair, J.~R., \& Cole, R.~H. 2015, Fund. Theor. Phys., 179, 783,
  \dodoi{10.1007/978-3-319-18335-0_23}

\bibitem[{Baibhav {et~al.}(2023)Baibhav, Cheung, Berti, Cardoso, Carullo,
  Cotesta, Del~Pozzo, \& Duque}]{Baibhav:2023clw}
Baibhav, V., Cheung, M. H.-Y., Berti, E., {et~al.} 2023, Phys. Rev. D, 108,
  104020, \dodoi{10.1103/PhysRevD.108.104020}

\bibitem[{Berti {et~al.}(2007)Berti, Cardoso, Cardoso, \&
  Cavaglia}]{Berti:2007zu}
Berti, E., Cardoso, J., Cardoso, V., \& Cavaglia, M. 2007, Phys. Rev. D, 76,
  104044, \dodoi{10.1103/PhysRevD.76.104044}

\bibitem[{Berti {et~al.}(2009)Berti, Cardoso, \& Starinets}]{Berti:2009kk}
Berti, E., Cardoso, V., \& Starinets, A.~O. 2009, Class. Quant. Grav., 26,
  163001, \dodoi{10.1088/0264-9381/26/16/163001}

\bibitem[{Berti {et~al.}(2006)Berti, Cardoso, \& Will}]{Berti:2005ys}
Berti, E., Cardoso, V., \& Will, C.~M. 2006, Phys. Rev. D, 73, 064030,
  \dodoi{10.1103/PhysRevD.73.064030}

\bibitem[{Berti {et~al.}(2016)Berti, Sesana, Barausse, Cardoso, \&
  Belczynski}]{Berti:2016lat}
Berti, E., Sesana, A., Barausse, E., Cardoso, V., \& Belczynski, K. 2016, Phys.
  Rev. Lett., 117, 101102, \dodoi{10.1103/PhysRevLett.117.101102}

\bibitem[{Bhagwat {et~al.}(2020)Bhagwat, Forteza, Pani, \&
  Ferrari}]{Bhagwat:2019dtm}
Bhagwat, S., Forteza, X.~J., Pani, P., \& Ferrari, V. 2020, Phys. Rev. D, 101,
  044033, \dodoi{10.1103/PhysRevD.101.044033}

\bibitem[{Boyle {et~al.}(2019)}]{Boyle:2019kee}
Boyle, M., {et~al.} 2019, Class. Quant. Grav., 36, 195006,
  \dodoi{10.1088/1361-6382/ab34e2}

\bibitem[{Brito {et~al.}(2018)Brito, Buonanno, \& Raymond}]{Brito:2018rfr}
Brito, R., Buonanno, A., \& Raymond, V. 2018, Phys. Rev. D, 98, 084038,
  \dodoi{10.1103/PhysRevD.98.084038}

\bibitem[{{Br{\"u}gmann} {et~al.}(2008){Br{\"u}gmann}, {Gonz{\'a}lez},
  {Hannam}, {Husa}, {Sperhake}, \& {Tichy}}]{brugmann_PRD2008}
{Br{\"u}gmann}, B., {Gonz{\'a}lez}, J.~A., {Hannam}, M., {et~al.} 2008, \prd,
  77, 024027, \dodoi{10.1103/PhysRevD.77.024027}

\bibitem[{{Bustillo} {et~al.}(2021){Bustillo}, {Lasky}, \&
  {Thrane}}]{2021PhRvD.103b4041B}
{Bustillo}, J.~C., {Lasky}, P.~D., \& {Thrane}, E. 2021, \prd, 103, 024041,
  \dodoi{10.1103/PhysRevD.103.024041}

\bibitem[{Cabero {et~al.}(2020)Cabero, Westerweck, Capano, Kumar, Nielsen, \&
  Krishnan}]{Cabero:2019zyt}
Cabero, M., Westerweck, J., Capano, C.~D., {et~al.} 2020, Phys. Rev. D, 101,
  064044, \dodoi{10.1103/PhysRevD.101.064044}

\bibitem[{Capano {et~al.}(2022)Capano, Abedi, Kastha, Nitz, Westerweck, Wang,
  Cabero, Nielsen, \& Krishnan}]{Capano:2022zqm}
Capano, C.~D., Abedi, J., Kastha, S., {et~al.} 2022, None.
\newblock \doarXiv{2209.00640}

\bibitem[{Capano {et~al.}(2023)Capano, Cabero, Westerweck, Abedi, Kastha, Nitz,
  Wang, Nielsen, \& Krishnan}]{Capano:2021etf}
Capano, C.~D., Cabero, M., Westerweck, J., {et~al.} 2023, Phys. Rev. Lett.,
  131, 221402, \dodoi{10.1103/PhysRevLett.131.221402}

\bibitem[{Carullo {et~al.}(2023)Carullo, Cotesta, Berti, \&
  Cardoso}]{Carullo:2023gtf}
Carullo, G., Cotesta, R., Berti, E., \& Cardoso, V. 2023, Phys. Rev. Lett.,
  131, 169002, \dodoi{10.1103/PhysRevLett.131.169002}

\bibitem[{Cheung {et~al.}(2021)Cheung, Poon, Chung, \& Li}]{Cheung:2020dxo}
Cheung, M. H.-Y., Poon, L. W.-H., Chung, A. K.-W., \& Li, T. G.~F. 2021, JCAP,
  02, 040, \dodoi{10.1088/1475-7516/2021/02/040}

\bibitem[{Cutler \& Schutz(2005)}]{Cutler:2005hc}
Cutler, C., \& Schutz, B.~F. 2005, Phys. Rev. D, 72, 063006,
  \dodoi{10.1103/PhysRevD.72.063006}

\bibitem[{Dreissigacker {et~al.}(2018)Dreissigacker, Prix, \&
  Wette}]{Dreissigacker:2018afk}
Dreissigacker, C., Prix, R., \& Wette, K. 2018, Phys. Rev. D, 98, 084058,
  \dodoi{10.1103/PhysRevD.98.084058}

\bibitem[{Dreyer {et~al.}(2004)Dreyer, Kelly, Krishnan, Finn, Garrison, \&
  Lopez-Aleman}]{Dreyer:2003bv}
Dreyer, O., Kelly, B.~J., Krishnan, B., {et~al.} 2004, Class. Quant. Grav., 21,
  787, \dodoi{10.1088/0264-9381/21/4/003}

\bibitem[{Estell\'es {et~al.}(2022)}]{Estelles:2021jnz}
Estell\'es, H., {et~al.} 2022, Astrophys. J., 924, 79,
  \dodoi{10.3847/1538-4357/ac33a0}

\bibitem[{{Giesler} {et~al.}(2019){Giesler}, {Isi}, {Scheel}, \&
  {Teukolsky}}]{Overtone_PRX_Giesler2019}
{Giesler}, M., {Isi}, M., {Scheel}, M.~A., \& {Teukolsky}, S.~A. 2019, Phys.
  Rev. X, 9, 041060, \dodoi{10.1103/PhysRevX.9.041060}

\bibitem[{Gossan {et~al.}(2012)Gossan, Veitch, \&
  Sathyaprakash}]{Gossan:2011ha}
Gossan, S., Veitch, J., \& Sathyaprakash, B.~S. 2012, Phys. Rev. D, 85, 124056,
  \dodoi{10.1103/PhysRevD.85.124056}

\bibitem[{Hild {et~al.}(2011)}]{Hild:2010id}
Hild, S., {et~al.} 2011, Class. Quant. Grav., 28, 094013,
  \dodoi{10.1088/0264-9381/28/9/094013}

\bibitem[{Hu \& Wu(2017)}]{1093nsrnwx116}
Hu, W.-R., \& Wu, Y.-L. 2017, National Science Review, 4, 685,
  \dodoi{10.1093/nsr/nwx116}

\bibitem[{Hunt(2019)}]{Hunt2019}
Hunt, J. 2019, Multiprocessing (Cham: Springer International Publishing),
  363--376, \dodoi{10.1007/978-3-030-25943-3_31}

\bibitem[{{Isi} \& {Farr}(2021)}]{2021arXiv210705609I}
{Isi}, M., \& {Farr}, W.~M. 2021, arXiv e-prints, arXiv:2107.05609.
\newblock \doarXiv{2107.05609}

\bibitem[{Isi \& Farr(2023)}]{Isi:2023nif}
Isi, M., \& Farr, W.~M. 2023, Phys. Rev. Lett., 131, 169001,
  \dodoi{10.1103/PhysRevLett.131.169001}

\bibitem[{Isi {et~al.}(2021)Isi, Farr, Giesler, Scheel, \&
  Teukolsky}]{Isi:2020tac}
Isi, M., Farr, W.~M., Giesler, M., Scheel, M.~A., \& Teukolsky, S.~A. 2021,
  Phys. Rev. Lett., 127, 011103, \dodoi{10.1103/PhysRevLett.127.011103}

\bibitem[{Isi {et~al.}(2019)Isi, Giesler, Farr, Scheel, \&
  Teukolsky}]{Isi:2019aib}
Isi, M., Giesler, M., Farr, W.~M., Scheel, M.~A., \& Teukolsky, S.~A. 2019,
  Phys. Rev. Lett., 123, 111102, \dodoi{10.1103/PhysRevLett.123.111102}

\bibitem[{Jaranowski {et~al.}(1998)Jaranowski, Krolak, \&
  Schutz}]{Jaranowski:1998qm}
Jaranowski, P., Krolak, A., \& Schutz, B.~F. 1998, Phys. Rev. D, 58, 063001,
  \dodoi{10.1103/PhysRevD.58.063001}

\bibitem[{Keppel(2012)}]{Keppel:2012ye}
Keppel, D. 2012, Phys. Rev. D, 86, 123010, \dodoi{10.1103/PhysRevD.86.123010}

\bibitem[{Luo {et~al.}(2016)}]{TQ_2015}
Luo, J., {et~al.} 2016, Class. Quant. Grav., 33, 035010,
  \dodoi{10.1088/0264-9381/33/3/035010}

\bibitem[{Ma {et~al.}(2023)Ma, Sun, \& Chen}]{Ma:2023cwe}
Ma, S., Sun, L., \& Chen, Y. 2023, Phys. Rev. Lett., 130, 141401,
  \dodoi{10.1103/PhysRevLett.130.141401}

\bibitem[{Maselli {et~al.}(2020)Maselli, Pani, Gualtieri, \&
  Berti}]{Maselli:2019mjd}
Maselli, A., Pani, P., Gualtieri, L., \& Berti, E. 2020, Phys. Rev. D, 101,
  024043, \dodoi{10.1103/PhysRevD.101.024043}

\bibitem[{Mei {et~al.}(2021)}]{TianQin:2020hid}
Mei, J., {et~al.} 2021, PTEP, 2021, 05A107, \dodoi{10.1093/ptep/ptaa114}

\bibitem[{Mishra {et~al.}(2022)Mishra, Ghosh, \& Chakraborty}]{Mishra:2021waw}
Mishra, A.~K., Ghosh, A., \& Chakraborty, S. 2022, Eur. Phys. J. C, 82, 820,
  \dodoi{10.1140/epjc/s10052-022-10788-x}

\bibitem[{Nee {et~al.}(2023)Nee, V\"olkel, \& Pfeiffer}]{Nee:2023osy}
Nee, P.~J., V\"olkel, S.~H., \& Pfeiffer, H.~P. 2023, Phys. Rev. D, 108,
  044032, \dodoi{10.1103/PhysRevD.108.044032}

\bibitem[{Nitz \& Capano(2021)}]{Nitz:2020mga}
Nitz, A.~H., \& Capano, C.~D. 2021, Astrophys. J. Lett., 907, L9,
  \dodoi{10.3847/2041-8213/abccc5}

\bibitem[{{Press}(1971)}]{GW_APJL_Press1971}
{Press}, W.~H. 1971, Astrophys. J. Lett., 170, L105, \dodoi{10.1086/180849}

\bibitem[{Prix \& Krishnan(2009)}]{Prix:2009tq}
Prix, R., \& Krishnan, B. 2009, Class. Quant. Grav., 26, 204013,
  \dodoi{10.1088/0264-9381/26/20/204013}

\bibitem[{{Punturo} {et~al.}(2010){Punturo}, {Abernathy},
  {et~al.}}]{2010CQGra..27s4002P}
{Punturo}, M., {Abernathy}, M., {et~al.} 2010, Class. Quantum Grav., 27,
  194002, \dodoi{10.1088/0264-9381/27/19/194002}

\bibitem[{{Reitze} {et~al.}(2019){Reitze}, {Adhikari},
  {et~al.}}]{2019BAAS...51g..35R}
{Reitze}, D., {Adhikari}, R.~X., {et~al.} 2019, in Bulletin of the American
  Astronomical Society, Vol.~51, 35.
\newblock \doarXiv{1907.04833}

\bibitem[{{Searle} {et~al.}(2009){Searle}, {Sutton}, \&
  {Tinto}}]{searleetal2009}
{Searle}, A.~C., {Sutton}, P.~J., \& {Tinto}, M. 2009, Classical and Quantum
  Gravity, 26, 155017, \dodoi{10.1088/0264-9381/26/15/155017}

\bibitem[{{Searle} {et~al.}(2008){Searle}, {Sutton}, {Tinto}, \&
  {Woan}}]{searleetal2008}
{Searle}, A.~C., {Sutton}, P.~J., {Tinto}, M., \& {Woan}, G. 2008, Classical
  and Quantum Gravity, 25, 114038, \dodoi{10.1088/0264-9381/25/11/114038}

\bibitem[{Siegel {et~al.}(2023)Siegel, Isi, \& Farr}]{Siegel:2023lxl}
Siegel, H., Isi, M., \& Farr, W.~M. 2023, Phys. Rev. D, 108, 064008,
  \dodoi{10.1103/PhysRevD.108.064008}

\bibitem[{Sieniawska \& Bejger(2019)}]{Sieniawska:2019hmd}
Sieniawska, M., \& Bejger, M. 2019, Universe, 5, 217,
  \dodoi{10.3390/universe5110217}

\bibitem[{{Speagle}(2020)}]{Dynesty_MNRAS_Speagle2020}
{Speagle}, J.~S. 2020, \mnras, 493, 3132, \dodoi{10.1093/mnras/staa278}

\bibitem[{Steltner {et~al.}(2023)Steltner, Papa, Eggenstein, Prix, Bensch,
  Allen, \& Machenschalk}]{Steltner:2023cfk}
Steltner, B., Papa, M.~A., Eggenstein, H.~B., {et~al.} 2023, Astrophys. J.,
  952, 55, \dodoi{10.3847/1538-4357/acdad4}

\bibitem[{{Teukolsky}(1973)}]{QNM_APJ_Teukolsky1973}
{Teukolsky}, S.~A. 1973, Astrophys. J., 185, 635, \dodoi{10.1086/152444}

\bibitem[{{Vishveshwara}(1970)}]{Schw_PRD_Vishveshwara1970}
{Vishveshwara}, C.~V. 1970, \prd, 1, 2870, \dodoi{10.1103/PhysRevD.1.2870}

\bibitem[{Wang \& Shao(2023)}]{Wang:2023mst}
Wang, H.-T., \& Shao, L. 2023, Phys. Rev. D, 108, 123018,
  \dodoi{10.1103/PhysRevD.108.123018}

\bibitem[{Wang \& Shao(2024)}]{Wang:2024liy}
---. 2024, Phys. Rev. D, 109, 043027, \dodoi{10.1103/PhysRevD.109.043027}

\bibitem[{{Wang} {et~al.}(2021){Wang}, {Tang}, {Li}, \&
  {Fan}}]{2021PhRvD.104j4063W}
{Wang}, H.-T., {Tang}, S.-P., {Li}, P.-C., \& {Fan}, Y.-Z. 2021, \prd, 104,
  104063, \dodoi{10.1103/PhysRevD.104.104063}

\bibitem[{Wang {et~al.}(2012)Wang, Shang, \& Babak}]{Wang:2012xh}
Wang, Y., Shang, Y., \& Babak, S. 2012, Phys. Rev. D, 86, 104050,
  \dodoi{10.1103/PhysRevD.86.104050}

\bibitem[{Welch(1967)}]{1967D.Welch}
Welch, P.~D. 1967, IEEE Trans. Audio \& Electroacoust, 15,
  \dodoi{10.1109/TAU.1967.1161901}

\bibitem[{Wette(2023)}]{Wette:2023dom}
Wette, K. 2023, Astropart. Phys., 153, 102880,
  \dodoi{10.1016/j.astropartphys.2023.102880}

\bibitem[{Yang {et~al.}(2017)Yang, Yagi, Blackman, Lehner, Paschalidis,
  Pretorius, \& Yunes}]{Yang:2017zxs}
Yang, H., Yagi, K., Blackman, J., {et~al.} 2017, Phys. Rev. Lett., 118, 161101,
  \dodoi{10.1103/PhysRevLett.118.161101}

\bibitem[{Zhu {et~al.}(2024)Zhu, Ripley, C\'ardenas-Avenda\~no, \&
  Pretorius}]{Zhu:2023mzv}
Zhu, H., Ripley, J.~L., C\'ardenas-Avenda\~no, A., \& Pretorius, F. 2024, Phys.
  Rev. D, 109, 044010, \dodoi{10.1103/PhysRevD.109.044010}

\end{thebibliography}

\end{document}